\documentclass[12pt]{article}
\usepackage{latexsym}
\usepackage{amsthm}
\usepackage{amsmath}
\usepackage{graphicx}
\usepackage{amssymb}
\usepackage{epsfig}
\usepackage{epstopdf}
\newcommand{\ra}{\ensuremath{\rangle}}
\newcommand{\ap}[1]{\langle #1\rangle}

\newcommand{\rr}[1]{\ensuremath{|#1\rangle}}

\newcommand{\be}{\begin{equation}}
\newcommand{\ee}{\end{equation}}
\newcommand{\bea}{\begin{eqnarray}}
\newcommand{\eea}{\end{eqnarray}}
\newcommand{\f}[2]{\ensuremath{\frac{#1}{#2}}}
\newcommand{\fap}[2]{\ensuremath{\frac{\ap{#1}}{\ap{#2}}}}
\newcommand{\fsp}[2]{\ensuremath{\frac{[#1]}{[#2]}}}
\newcommand{\sh}[1]{\ensuremath{[#1\rangle}}
\newcommand{\toZ}{\ensuremath{\rightarrow 0}}

\newcommand{\bef}{\begin{figure}[htbp]\begin{center}}
\newcommand{\eef}{\end{center}\end{figure}}
\begin{document}
\baselineskip=18pt
\setcounter{footnote}{0}
\setcounter{figure}{0} \setcounter{table}{0}
%%%%%%%%%%%%%%%%%%%%%%%%%%%%%%%%%%%%%%%%%%
\begin{titlepage}
\begin{flushright}
SLAC-PUB-13426\qquad \\
SU-ITP-08/32\qquad\ \\
\end{flushright}
\vspace{.0in}

\begin{center}
\vspace{1cm}

{\Large \bf Constructing the Tree-Level Yang-Mills S-matrix Using Complex Factorization}

\vspace{0.8cm}

{\bf Philip C. Schuster$^1$, Natalia Toro$^2$}

\vspace{0.5cm}
{\it $^1$ Theory Group, SLAC National Accelerator Laboratory, \\
Menlo Park, CA 94025, USA}

\vspace{0.2cm}
{\it $^2$ Stanford Institute for Theoretical Physics, Stanford University, \\
Stanford, CA 94305, USA}

\end{center}
\vspace{1cm}

\begin{abstract}
A remarkable connection between BCFW recursion relations and
constraints on the S-matrix was made by Benincasa and Cachazo in
0705.4305, who noted that mutual consistency of different BCFW
constructions of four-particle amplitudes generates
non-trivial (but familiar) constraints on three-particle coupling
constants --- these include gauge invariance, the equivalence principle, and
the lack of non-trivial couplings for spins $>2$. These constraints
can also be derived with weaker assumptions, by demanding the
existence of four-point amplitudes that factorize properly in all
unitarity limits with complex momenta. From this starting point, we
show that the BCFW prescription can be interpreted as an algorithm for
{\it fully} constructing a tree-level S-matrix, and that complex
factorization of general BCFW amplitudes follows from the
factorization of four-particle amplitudes.  The allowed set of BCFW
deformations is identified, formulated entirely as a statement on the
three-particle sector, and using {\it only} complex factorization as a
guide. Consequently, our analysis based on the physical
consistency of the S-matrix is entirely independent of field
theory. We analyze the case of pure Yang-Mills, and outline a proof
for gravity. For Yang-Mills, we also show that the well-known
scaling behavior of BCFW-deformed amplitudes at large z is a simple
consequence of factorization. For gravity, factorization in certain
channels requires asymptotic behavior $\sim 1/z^2$.
\end{abstract}

%%\bigskip
%%\bigskip

\end{titlepage}
%%%%%%%%%%%%%%%%%%%%%%%%%%%%%%%%%%%%%%%%%%%
%\tableofcontents \vfill\eject
%%%%%%%%%%%%%%%%%%%%%%%%%%%%%%%%%%%%%%%%%%%
%%%%%%%%%%%%%%%%%%%%%%%%%%%%%%%%%%%%%%%%%%%
\section{Introduction}
Gauge theories represent a highly constrained framework for
describing interacting spin-1 particles.  Weinberg's seminal papers of
1964-65 \cite{Weinberg:1964ev,Weinberg:1964ew,Weinberg:1965rz} demonstrated that many of these constraints
can be seen as inevitable consequences of requiring scattering
amplitudes to be both unitary and Lorentz-invariant.  In particular,
both charge conservation and Maxwell's equations follow from S-matrix
arguments alone.  Likewise, consistency of the spin-2 S-matrix
requires the equivalence principle and Einstein's equations at
tree-level.

A concrete and beautiful confirmation that much of the structure of
gauge and gravity theories is contained in their S-matrix is the
existence of purely on-shell recursion relations for gauge
theory \cite{Britto:2004ap,Britto:2005fq} and gravity
\cite{Bedford:2005yy,Cachazo:2005ca,Benincasa:2007qj,ArkaniHamed:2008yf}, which allow
the calculation of on-shell scattering amplitudes entirely in terms of
lower-point on-shell amplitudes.  However, these relations have always been
derived from the underlying local field theories; their relation to S-matrix consistency 
arguments such as \cite{Weinberg:1964ev,Weinberg:1964ew,Weinberg:1965rz} were unclear.  

A striking connection between BCFW recursion and consistency of the
spin-1 and spin-2 S-matrix was uncovered by Benincasa and
Cachazo \cite{Benincasa:2007xk}.  These authors introduced a ``four-particle
test''--- the requirement that two BCFW shifts generate the same answer
for any four-point amplitude.  This requirement could only be met when
the ``coupling constants'' (coefficients of three-point amplitudes)
satisfied non-trivial relations.  For instance, it was shown
in \cite{Benincasa:2007xk} that the four-particle test generates the Jacobi identity
for spin-1 particles.  However, the four-particle test only makes
sense when each of the BCFW constructions involved is valid (i.e. field
theory amplitudes vanish in the limit of large shifts).  The
constraints obtained are reminiscent of those identified by Weinberg
as consequences of Lorentz invariance and unitarity.  The physical origin
of the constraints obtained in \cite{Benincasa:2007xk} is somewhat
unclear, however, as the argument assumes the validity of BCFW
constructions.  An open question in 
\cite{Benincasa:2007xk} was whether new constraints would be found by
applying the same criterion to higher-point amplitudes.  This question
is also considered in \cite{SongHe}.

After a brief review of the spinor-helicity formalism and BCFW
recursion in Section \ref{sec:review}, we show in Section
\ref{sec:consistency} that the conditions from the four-particle test
can be understood as the result of demanding Lorentz invariance and an
analytic continuation of unitarity to complex momenta, ``complex
factorization''.  The latter is simply the requirement that amplitudes
factorize into sub-amplitudes when an intermediate complex momentum
can go on-shell.  A byproduct of this treatment is a simple criterion
for BCFW shifts --- shifts of the $\sh{+-}$ type (the only invalid shift in
gauge theories) cannot possibly satisfy complex factorization in all
limits.

BCFW recursion relations provide a formula for generating a set of
arbitrary high-point amplitudes for spin-1 massless interacting
particles.  In Section \ref{sec:4} we show that amplitudes
generated by BCFW recursion are guaranteed to satisfy complex
factorization, provided the four-particle consistency 
requirements on couplings and shifts are satisfied.  We also provide
an S-matrix derivation  of the large-$z$ scaling of BCFW amplitudes
(as $z^3$ for $\sh{+-}$ deformations and $1/z$ for all others), which
is used in our factorization argument.
Thus, we have
taken the programs described above fully on-shell: conditions on the
structure of spin-1 amplitudes, and a construction of higher-point
amplitudes from lower-point ones are justified with no reference
to a local gauge theory.

We also outline the analogous result for gravity, and highlight the
differences.  One piece of the proof is missing: the argument relies on the
scaling behavior of $n$-point spin-2 amplitudes at large BCF shift
parameters $z$ (i.e. growing as $z^6$ for $\sh{+-}$ shifts, and
falling as $z^{-2}$ for
all others); an on-shell proof of this scaling result would complete the proof
of factorization for BCFW gravity amplitudes.

\section{Review of Formalism and Notation}\label{sec:review}
In this section, we review two essential elements of the constructions
in the remainder of this paper: the spinor-helicity formalism \cite{Witten:2003nn,Berends:1981rb,DeCausmaecker:1981bg,Kleiss:1985yh} and BCFW
recursion relations \cite{Britto:2005fq,Britto:2004ap,Dixon:1996wi,Bern:2007dw}.  The spinor-helicity formalism will be useful as
a means of writing down manifestly Lorentz-invariant amplitudes for
higher-spin massless particles, involving only the physical
interacting degrees of freedom.  BCFW recursion will be used as a
means of generating amplitudes that satisfy unitarity in a subset of
poles; however, the amplitudes thus constructed are not manifestly unitary in \emph{all} limits ---
demanding unitarity in all remaining channels will give constraints on the fundamental 
three-particle couplings and on the set of allowed BCFW shifts, as
discussed in Section \ref{sec:consistency}.

\subsection{Spinor-Helicity Formalism and Three-Point Amplitudes}
Any four-vector can be related to a bispinor by $p_{\alpha
  \dot\alpha}= \sigma^\mu_{\alpha\dot\alpha} p_\mu$; when $p^2=0$,
$p_{\alpha \dot\alpha}$ has rank one, and we can write $p_{\alpha
  \dot\alpha}=\lambda_{\alpha}\tilde\lambda_{\dot\alpha}$.  The
spinors $\lambda_\alpha$ and $\tilde \lambda_{\dot\alpha}$ are
uniquely determined by $p_\mu$, up to a complex rescaling
$(\lambda,\tilde \lambda) \rightarrow (z \lambda, \f{1}{z} \tilde
\lambda)$, and they transform in the $(1/2,0)$ and $(0,1/2)$
representations of the Lorentz group; they also transform simply under
helicity rotations, $(\lambda,\tilde \lambda) \rightarrow (e^{-i\theta/2}
\lambda, e^{+i\theta/2} \tilde \lambda)$.  In this notation, two-particle momentum
invariants can be written as 
\be
p_i.p_j = \ap{ij}[ij];
\ee
momentum conservation is the bi-spinor condition 
\be
\sum_{i}{\rr{i}|i]} = 0,
\ee
and the Schouten identity
\be
[ij]|k] + [jk]|i] + [ki]|j] = 0
\ee
follows from the antisymmetry of spinor products and the
fact that spinors live in a two-dimensional vector space.

The transformation properties of $\lambda$,$\tilde \lambda$ under
both Lorentz transformations and little group helicity rotations make
them very useful for  formulating an on-shell theory 
of spin-s particles --- all of the Lorentz transformation properties of states are neatly encoded 
in the spinors.  The helicity rotation operator associated with any
external momentum $i$ is
\be
H_i = -\f{1}{2} \lambda \f{d}{d\lambda} + \f{1}{2}\tilde\lambda
\f{d}{d\tilde\lambda};
\label{eqn:helOperator}
\ee
Lorentz-invariant of an $n$-point amplitude in which the $i$'th particle has helicity
$h_i$ is guaranteed if 
\be
H_i A = h_i A
\ee 
for all legs, and $A$ has no free
spinor indices.  Amplitudes are then naturally expressed in
terms of Lorentz invariant ``holomorphic'' spinor products $\ap{\lambda\mu}=\epsilon_{\alpha\beta}\lambda^{\alpha}\mu^{\beta}$ and
 ``anti-holomorphic'' products $[\tilde\lambda\tilde\mu]=\epsilon_{\dot\alpha\dot\beta}\tilde\lambda^{\dot\alpha}\tilde\mu^{\dot\beta}$. 

In fact, demanding the helicity transformation properties above fixes
three-particle amplitudes completely up to coupling coefficients
\cite{Benincasa:2007xk}.  Note that the on-shell condition for three
particles can be satisfied in two ways, either
$[12]=[13]=[23]=0$ or $\ap{12}=\ap{13}=\ap{23}=0$; though all real
momentum invariants vanish in this limit, the spinor products
$\ap{ij}$ can be non-zero in the first case, as can $[ij]$ in the
latter case.  
For example, a theory of several interacting massless spin-s 
$(s\geq 1)$ particles has three-particle amplitudes (fixed by Lorentz invariance),
\begin{eqnarray}
A_3^{(a)}(i^{+,\alpha_i},j^{+,\alpha_j},k^{-,\alpha_k}) &=& f^{(a)}_{\alpha_i\alpha_j\alpha_k}\left(\frac{[ij]^3}{[jk][ki]}\right)^s,  \nonumber \\
A_3^{(h)}(i^{-,\alpha_i},j^{-,\alpha_j},k^{+,\alpha_k}) &=& f^{(h)}_{\alpha_i\alpha_j\alpha_k}\left(\frac{\ap{ij}^3}{\ap{jk}\ap{ki}}\right)^s, \label{vectors}
\end{eqnarray}
where $+,-$ refers to helicity, indexes $\alpha_i,\alpha_j,\alpha_k$
label species of particles, and $i,j,k$ label four-momentum
spinors. Demanding the amplitudes have crossing symmetry and are
invariant under exchange of states requires the
$f^{(a,h)}_{\alpha\beta\gamma}$ be completely anti-symmetric
\cite{Benincasa:2007xk}. Under parity, $+$ and $-$ are exchanged. Demanding that pure
spin-s interactions be invariant under parity requires
$f^{(a)}=f^{(h)}$, an assumption that we'll make throughout this
paper. To minimize the complexity of notation, we will often use the
spinor labels $i,j,k$ to refer to species labels.  There is another
set of three-particle amplitudes $A^{(a)}(i^+,j^+,k^+)$ and
$A^{(h)}(i^-,j^-,k^-)$ consistent with Lorentz invariance; however we
will assume their coefficients are zero.  

For later use, we also note the three-particle amplitudes for one spin-s particles with a collection of scalars:
\begin{eqnarray}
A_3(i^{a_i},j^{a_j};k^{+,\alpha_k}) &=& \kappa_{a_ia_j}^{\alpha_k}\left(\frac{[ki][jk]}{[ij]}\right)^s,  \nonumber \\
A_3(i^{a_i},j^{a_j};k^{-,\alpha_k}) &=& \kappa_{a_ia_j}^{\alpha_k}\left(\frac{\ap{ki}\ap{jk}}{\ap{ij}}\right)^s, \label{vectorscalar}
\end{eqnarray}
where we have again assumed a parity, and $a_i$ labels species of scalars. In this case as well, the $k_{ij}^a$ should be anti-symmetric in $i,j$ in order for the amplitudes to be symmetric under interchange of scalars. 

%% In this paper, we'll write all amplitudes in the spinor-helicity language. Nonetheless, should one want to 
%% use the clumsier language of four-vectors and polarization tensors, we can re-introduce gauge freedom 
%% to the description and write down polarization tensors by choosing reference spinors $\rr{\mu}$ and $|\mu]$:
%% \be
%% \epsilon^+_\mu = \sigma_{\mu,\alpha\dot\alpha} \frac{
%% \mu^{\alpha} \tilde \lambda^{\dot\alpha} }{\lambda\dot \mu} = \frac{\rr{\mu}|\tilde\lambda]}{\ap{\lambda\mu}}
%% \qquad
%% \epsilon^-_\mu = \sigma_{\mu,\alpha\dot\alpha} \frac{
%% \lambda^{\alpha} \tilde \mu^{\dot\alpha}}{\tilde\lambda\dot \tilde\mu}
%% =\frac{\rr{\lambda}|\tilde\mu]}{[\lambda\mu]};
%% \ee
%% and we see that these have the proper helicity transformations. The
%% statement of gauge-invariance is that amplitudes do not depend on the
%% arbitrarily chosen spinors $\rr{\mu}$ and $|\tilde \mu]$.  

\subsection{BCFW Recursion}\label{sec:BCF}
The BCFW formalism \cite{Britto:2005fq,Britto:2004ap,Dixon:1996wi,Bern:2007dw} can be expressed very concisely in spinor-helicity formalism, but we first summarize it directly in four-momentum space. Consider the amplitude $A(p_1, h_1, p_2,h_2, \dots, p_n, h_n)$, where $h_i$ labels helicity.  The basic idea of the BCFW formalism is to deform amplitudes into a function of a single complex variable $z$, and then re-express the amplitude in terms of residues. 

The simplest complex deformation of the amplitude that keeps all momenta on-shell is a deformation involving only two external legs. Consider the legs $p_1$ and $p_2$. Choose an arbitrary null four-vector $q$ such that $q.p_1=q.p_2=0$. Then we can deform by,
\be
\hat p_1(z) = p_1 + z q,\, \hat p_2(z) = p_2 - z q,
\ee
which keeps $p_1(z)$ and $p_2(z)$ null. In the spinor-helicity
language, this is satisfied by this isimplemented particularly easy to
implement by $q = \rr{1}|2]$, so that the BCFW shift only deforms one
  of the two spinors associated with each leg:
\be |\hat 1](z) \equiv |1] + z |2], \quad |\hat 2\ra(z) = \rr{2} - z
  \rr{1},\quad(\rr{1},\,|2]\mbox{ fixed}) \ee
We will call this the $\sh{1,2}$ shift.  Since $q.p_1=q.p_2 = q^2=0$, any kinematic
invariant $(\sum p_{i})^2$ is at most linear in $z$.  These deformation naturally make the full on-shell amplitude a function of $z$,
\be
A(z) \equiv A(\hat p_1(z), h_1, \hat p_2(z),h_2, \dots, p_n, h_n).
\ee
At tree level, $A(z)$ is a rational function of $z$ and so is fully determined by its poles and behavior as $z\toZ$. 

If we assume tree-level factorization, the only poles in the amplitude arise from propagators going on-shell, and their residues are fully
determined by lower-point amplitudes.  If $A(z)\toZ$ as $z\toZ$, the amplitude is then fully determined by products of lower-point on-shell amplitudes. In this special case, we obtain the BCFW recursion relation \cite{Britto:2004ap,Britto:2005fq},
\be
A_{BCF}(1,2,\dots, N)  \equiv \sum_{L, R} \sum_{h_{K}}A(\hat 1, L, -\hat K^{-h_K})
\f{1}{K^2} A(\hat K^{h_K}, \hat 2, R),
\label{eqn:bcf12}
\ee
where the sum is over partitions of $3, 4, \dots, N$ into two sets $L$
and $R$, $K=\sum_{i\in\{1,L\}} p_i$ is the (generically non-null) momentum
flowing out of the right factor before the BCF shift, and $\hat K = \hat
p_1(z) + \sum_{i\in L}p_i$ is the null momentum flowing out of the right
graph \emph{after} the BCF shift.

This expression transforms properly under all helicity rotations
\eqref{eqn:helOperator} ($H_i A^{(n)} = h_i A^{n}$) so long as the
same is true of the lower-point amplitudes from which it is generated.
Thus, Lorentz invariance of BCFW amplitudes is manifest.   

\section{Consistency of Four-Point Amplitudes}\label{sec:consistency}
In this section, we study the structure of consistent four-point
amplitudes for four massless particles, satisfying two conditions ---
Lorentz invariance and a strong version of tree-level unitarity
\emph{in unconstrained complex momenta}, which we will refer to as
complex factorization.  We begin by explaining the conditions in some
detail, then build amplitudes for specific examples.  The general
pattern that emerges is that, for high-spin theories, Lorentz
invariance and factorization in a subset of channels fully constrains
the leading behavior of the amplitude (in the sense of
power-counting).  This amplitude will only be able to satisfy unitarity in the
remaining channel(s) if the coefficients of different three-point
amplitudes are related. \footnote{The authors thank N. Arkani-Hamed for
  suggesting this approach.}

Our results are closely related to those of \cite{Benincasa:2007xk}
(and very much motivated by that work), but we make a significantly
weaker set of assumptions.  Specifically, the authors of
\cite{Benincasa:2007xk} find conditions that three-point amplitudes
must satisfy \emph{if four-point amplitudes can be constructed by a
  BCFW recursion}.  Therefore, their argument relies on
field-theoretic derivations of the validity of different BCFW shifts.
Since our goal is to motivate the self-consistency of BCFW
constructions independent of field theory, and find an S-matrix
criterion for their validity, it is important that we \emph{do not}
assume this.  However, it is easy to see that when the assumptions of
\cite{Benincasa:2007xk} are satisfied, the two methods will give the
same consistency conditions on three-point amplitudes. 

\subsection{Setup and Interacting Spin-1}
We begin by setting up the constraints on four-particle scattering
amplitudes from Lorentz invariance, and explaining the requirement of
complex factorization.  We will consider the amplitude for scattering
of four spin-1 particles as an explicit example, and derive the Jacobi
identity.  

We first demand that the four-particle scattering amplitude be a
Lorentz scalar, and tranform as a product of one-particle states under
independent helicity rotations of each.  This condition is easily
imposed in the spinor-helicity formalism --- the only non-vanishing
scalar invariants (under Lorentz invariance \emph{and} individual
helicity rotations) are
\bea 
s=(p_1+p_2)^2 = \ap{12}[12]=\ap{34}[34],\\ 
t=(p_1+p_3)^2 = \ap{13}[13]=\ap{24}[24],\\ 
u=(p_1+p_4)^2 = \ap{14}[14]=\ap{23}[23] = -s -t.\\ 
\eea
If we define an arbitrary particular solution $\mathcal{H}(1,2,3,4)$
that transforms correctly under helicity rotations, then a general
four-point amplitude has the form
\be
A(1^{h_1}, 2^{h_2}, 3^{h_3}, 4^{h_4}) = \mathcal{H}(1,2,3,4) \; \times
\; f(s,t,u)
\ee
for some function $f$.  It will be most convenient to choose an
$\mathcal{H}$ that is polynomial in the spinor-product invariants, but does
not contain any Mandelstam scalar invariants.  For example, for four
spin-1 particles of helicities $+,+,-,-$, 
\be
\mathcal{H}(1-,2-,3+,4+) = \ap{12}^2 [34]^2.
\ee

The requirement of complex factorization is the familiar tree-level
unitarity --- when any sum of momenta in a diagram goes on-shell, it
gives rise to a single pole, associated with splitting the diagram in
two, e.g.  
\be 
\lim_{s\toZ } s\times A(1,2,3,4) = \sum_{h,a}
A(1,2,-P_{12}^{-h}) A(3,4,P_{12}^h),\label{unitarity} 
\ee 
where we sum over all allowed intermediate helicities, and a possible
species index $a$ (we will drop $a$ in much of the discussion).
We give this familiar criterion a new name, because we will require it
to hold at arbitrary complex momenta. We have seen already that for
particles of non-zero spin, there are two distinct three-point
amplitudes in different on-shell limits --- an $A^{(h)}(1,2,3)$ when
$[ij]=0$ and an $A^{(a)}(1,2,3)$ when $\ap{ij}=0$.  If $h_1 + h_2 +
h_3 \neq 0$, then both $A^{(a)}$ and $A^{(h)}$ vanish in the
real-momentum collinear limit, where $[ij]$ and $\ap{ij}$ both go to
zero.  This is a more general complexification of momenta than the
usual analytic continuation of the Mandelstam variables, and we will
see in a moment that there are cases where a real-momentum
$s\rightarrow 0$ limit is trivial, but one of the two complex
directions is not.  

Which three-point amplitudes appear in the unitarity condition
\ref{unitarity} depends on how we take the limit $s\rightarrow 0$.  We
can take either
\be
\lim_{[12],\ap{34}\toZ}  s\times A(1,2,3,4) = \sum_{h}
A^{(h)}(1,2,-P_{12}^{-h}) A^{(a)}(P_{12}^h,3,4)\label{schan}
\ee
or
\be
\lim_{\ap{12},[34]\toZ}  s\times A(1,2,3,4) = \sum_{h}
A^{(a)}(1,2,-P_{12}^{-h}) A^{(h)}(P_{12}^h,3,4).\label{sstar}
\ee
The limit $[12]\toZ$ and $[34]\toZ$ also enforces either $[ij]\toZ$
for all $i,j$ or a soft limit on one of the three legs; we will not
demand unitarity in this limit.  We can thus refer unambiguously to
the two limits above as $[12]\toZ$ \eqref{schan} or $\ap{12}\toZ$
\eqref{sstar}.  

Having stated the requirement of complex factorization, let us apply
it to the four-gauge-boson amplitude.  One way of obtaining an
expression that factorizes in the limits $[13]\toZ$ and $[14]\toZ$ is
by using the BCFW formula, using a shift
 $ |\hat 1](z) \equiv |1] + z |2]$, $\quad |\hat 2\ra(z) = \rr{2} -
      z\rr{1}$ (called a $\sh{12}$ shift).
The two terms in the formula are shown in Figure
\ref{fig:FourParticleConst}
\bef
\includegraphics[height=1.6in]{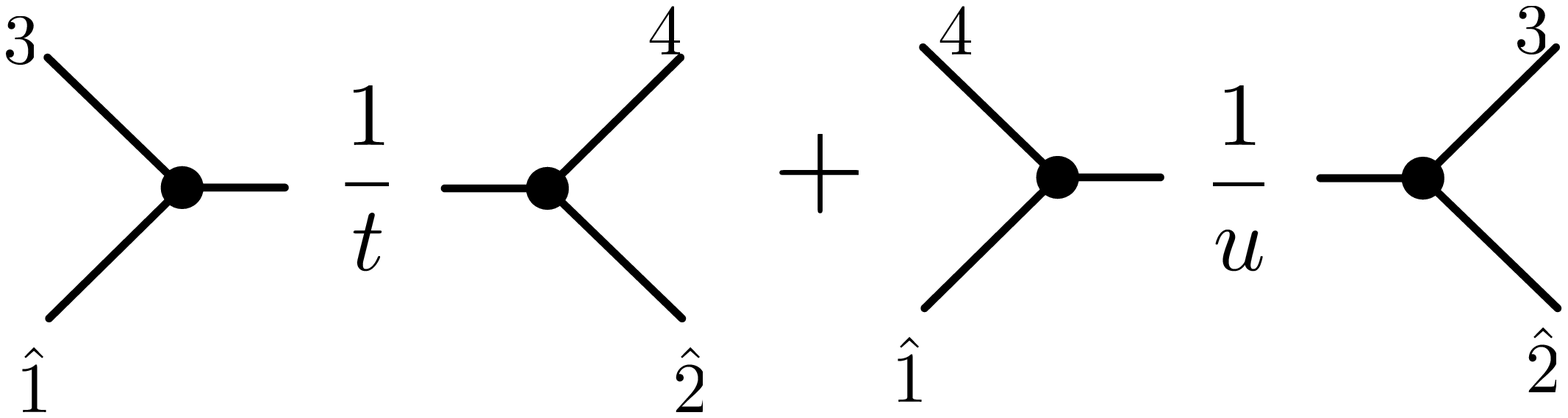}
\caption{The two diagrams in the BCFW construction of a four-point amplitude $A(1,2,3,4)$ using a $\sh{12}$ shift. Both $t$- and $u$-channel poles are exposed in the left and right pieces respectively.\label{fig:FourParticleConst}}
\eef
The result is
\be
A_{BCF,\sh{1,2}}(1^-,2^-,3^+,4^+)=\ap{12}^2[34]^2\left[\frac{f_{\beta13}f_{\beta42}}{st} - \frac{f_{\alpha14}f_{\alpha32}}{su}\right],\label{eqn:4partgauge}
\ee
where repeated indexes are summed.  The fact that we have obtained
this amplitude by a BCFW construction is quite incidental ---
it is the unique amplitude one can write down, involving only
dimensionless coupling constants, that factorizes in the limits
$[13]\toZ $ and $[14]\toZ$ and has no unphysical poles.  For example,
adding a term $\sim \frac{1}{(s+2t)(s+5t)}$ to the scalar function in
brackets would introduce two single poles at unphysical locations,
where no particle goes on-shell; adding a contribution $\sim 1/s^2$
would introduce an unphysical double singularity as $[12]\toZ$, and of
course adding a term $\sim \f{1}{st}$ or $\sim \f{1}{ut}$ would change
the $t$- and $u$-channel singularities.

One can check that the $\ap{13}$ and $\ap{14}\toZ$ singularities in
\eqref{eqn:4partgauge} have the proper form, and they do.  More
interesting conditions come from the $s$-channel.  We note that,
because of the helicity factor, \eqref{eqn:4partgauge} has a pole as
$[12]\toZ$ but not as $\ap{12}\toZ$.  In fact, the real
limit with $\ap{12},[12]$ vanishing simultaneously \emph{also} vanishes.
But the limit $[12]\toZ$ is non-zero.  Complex factorization
\eqref{schan}
requires
\begin{eqnarray}
\lim_{[12]\toZ}\ap{12}[12]A(1^-,2^-,3^+,4^+) &=& A^{(h)}(1^-,2^-,-P^{-h}_{12})A^{(a)}(P^{h}_{12},3^+,4^+) , \\
                                           &=& \ap{12}^2[34]^2\left(\frac{f_{12\alpha}f_{34\alpha}}{t}\right),
\end{eqnarray}
while $A_{BCF}$ above has
\be
\lim_{[12]\toZ}\ap{12}[12]A_{BCF}(1^-,2^-,3^+,4^+) = \ap{12}^2[34]^2\frac{1}{t} ( f_{\alpha13}f_{\alpha42} + f_{\alpha14}f_{\alpha32} ) 
\ee
For the two expressions to agree, the couplings must satisfy a Jacobi
Identity:
\be
f_{\alpha_1\alpha_4\beta}f_{\beta\alpha_3\alpha_2}+f_{\alpha_1\alpha_3\beta}f_{\beta\alpha_4\alpha_2}+f_{12\beta}f_{\beta34}=0
\ee 
(where we've restored the full species label for clarity). This result
was also found in \cite{Benincasa:2007xk}, by demanding agreement of
two BCFW shifts involving different legs.  It is clear why these give
the same result: the two BCFW shifts are imposing factorization in
different poles!

We have now obtained a four-point amplitude that can satisfy complex
factorization in all channels; it is unique up to less singular terms,
which by power-counting must have new, dimensionful couplings.
Writing down \emph{any} such amplitude required a relation between
three-point coefficients.  With this amplitude in hand, we can now
check BCFW recursion relations explicitly, with different shifts.  We
have already seen that the $\sh{1,2}$ shift gives the correct
amplitude; so do  the $\sh{1,3}$ shift and all shifts of legs with
helicities $\sh{--}$, $\sh{+-}$, or $\sh{++}$.  However, if we try to
build an amplitude with the $\sh{3+,1-}$ shift, we find
\be
A_{BCF;\sh{31}}(1^-,2^-,3^+,4^+) \propto \ap{12}^2 [34]^2
\frac{t^3}{s^4 u}.
\ee
This satisfies factorization in the $[32]/\ap{14}$  and $[34]/\ap{12}$
limits ($A \sim 1/{su}$ as $u\rightarrow 0$ and is non-singular as
$\ap{12}\toZ$), but has an unphysical fourth-power singularity as
$[12]\toZ$ and no $t$-channel singularities.  So we can see directly
that this shift cannot produce physical amplitudes.
Our finding that all shifts except $\sh{+-}$ are valid agrees with the
gauge-theory result \cite{Britto:2004ap,Britto:2005fq}.

\subsection{Spin-1 With Matter}
We can repeat this analysis for spin-1 interactions interacting with
massless matter (for simplicity, we consider scalars) labelled by
indexes $a_i$, with three-particle amplitudes
given by equation \ref{vectorscalar}. Consider the four-particle
amplitude $A(1^-,2^+,3,4)$. We again use a $\sh{12}$ shift to build
an amplitude, and again it is consistent with factorization in both
$t$-channel limits and both $u$-channel limits (though the
factorization as $\ap{1j}\toZ$ does not follow from the BCFW
construction, and must be checked explicitly):
\be
A_{BCF}(1^-,2^+,3,4) = \ap{13}\ap{14}[23][24] \left( \frac{\kappa_{3a}^1\kappa_{a4}^2}{st} + \frac{\kappa_{4a}^1\kappa_{a3}^2}{su} \right) .
\ee
Again, any modification with the same energy-scaling would violate
complex factorization in the $t$ or $u$-channels, or introduce
unphysical singularities elsewhere.  
As $[12]\toZ$, complex factorization requires
\begin{eqnarray}
\lim_{[12]\toZ}\ap{12}[12]A(1^-,2^+,3,4) &= A^{(h)}(1^-,2^+,K_{12})A^{(a)}(-K_{12},3,4) \\
							&= -\ap{13}\ap{14}[23][24] ( \frac{f_{\alpha12}\kappa_{34}^{\alpha}}{t} ),
\end{eqnarray}
and in this case the $\ap{12}\toZ$ limit is identical.
The limit of the BCFW construction is 
\begin{eqnarray}
\lim_{[12]\toZ}\ap{12}[12]A_{BCF}(1^-,2^+,3,4) &= \ap{13}\ap{14}[23][24] \frac{1}{t} ( \kappa_{3a}^1\kappa_{a4}^2 - \kappa_{4a}^1\kappa_{a3}^2 ),
\end{eqnarray}
which satisfies factorization if
\be
\kappa_{a_3b}^{\alpha_1}\kappa_{ba_4}^{\alpha_2} - \kappa_{a_3b}^{\alpha_2}\kappa_{ba_4}^{\alpha_1} = -f_{\beta\alpha_1\alpha_2}\kappa_{a_3a_4}^{\beta}
\ee 
(where we have again restored full species labels). We recognize this
as the requirement that the  $\kappa_{ij}^{\alpha}$ furnish a representation of the Lie
Algebra defined by the $f_{\alpha\beta\gamma}$.

For self-interacting scalars with a three-point amplitude 
$A(a_i,a_j,a_k)=\gamma_{a_ia_ia_k}$, we can apply $s$-channel
factorization to the amplitude $A(1^-,2,3,4)$ constructed with an
$\sh{12}$ shift, and find that the scalar interaction must satisfy
charge conservation. We do not show this explicitly here.  One could
also check using this amplitude or the previous one that at
four-point,  shifts $\sh{-,0}$ and $\sh{0,+}$ are valid but shifts
$\sh{+,0}$ or $\sh{0,-}$ are not, consistent with the general
field-theory result of \cite{Cheung:2008dn}.  

\subsection{Spin-2}

We can repeat the above analysis for massless interacting spin-2
particles. First consider the amplitude $A(1^-,2^-,3^+,4^+)$
constructed from a $\sh{12}$ shift ($+$ and $-$ labels $+2$ and $-2$
helicity states). Using the BCFW ansatz, we obtain,
\be
A_{BCF}(1^-,2^-,3^+,4^+) = [34]^4\ap{12}^4 \left( \frac{f_{13\alpha}f_{24\alpha}}{s^2t} + \frac{f_{14\alpha}f_{23\alpha}}{s^2u} \right).
\ee
Interestingly, each term separately has a double pole as $s\toZ$ (taking $[12]\toZ$). Complex factorization in this channel requires
\begin{eqnarray}
\lim_{[12]\toZ}\ap{12}[12]A_{BCF}(1^-,2^-,3^+,4^+) &=&
\lim_{[12]\toZ}\ap{12}^4[34]^4 \left( \frac{f_{13\alpha}f_{24\alpha}
  \cdot u
  + f_{14\alpha}f_{23\alpha} \cdot t}{stu} \right) , \nonumber \\
&=& \ap{12}^4[34]^4 \frac{f_{12\alpha}f_{34\alpha}}{t^2} .
\end{eqnarray}
To make the double pole vanish, we require
\be
f_{14\alpha}f_{23\alpha}=f_{13\alpha}f_{24\alpha},
\ee
(so that $t+u=-s$ in the numerator cancels an $s$ in the denominator), 
while getting the correct coefficient on the single pole requires
additionally
\be
f_{13\alpha}f_{24\alpha}=f_{12\alpha}f_{34\alpha}.
\ee
As pointed out in \cite{Benincasa:2007xk}, these relations imply that
the $f_{\alpha\beta\gamma}$ furnish an algebra that is commutative and
associative, and therefore any multi-graviton theory can be reduced to
a theory of self-interacting gravitons that decouple from one
another.  From here on, we will consider only one spin-2 species.

We next consider a single massless spin-2 particle coupled to a set of
interacting scalars. As before, the fundamental three-particle scalar
interactions are just constants
$A(a_i,a_j,a_k)=\gamma_{a_ia_ia_k}$. The fundamental spin-2-scalar
three-particle amplitudes are 
\begin{eqnarray}
A_3(i^{a_i},j^{a_i};k^+) &=& g_{a_i}(\frac{[ki][jk]}{[ij]})^2,  \nonumber \\
A_3(i^{a_i},j^{a_i};k^-) &=& g_{a_i}(\frac{\ap{ki}\ap{jk}}{\ap{ij}})^2,
\end{eqnarray}
(by symmetry, the spin-2-scalar couplings must be symmetric in the
$a_i$, so we have diagonalized them). We now consider building the amplitude $A(1^-,2,3,4)$ using a $\sh{12}$ shift. The BCFW construction gives, 
\be
A_{BCF}(1^-,2,3,4) = \ap{14}^2\ap{13}^2[34]^2 \gamma_{234} \left( \frac{g_{33}u+g_{44}t}{s^2tu} \right).
\ee
Demanding complex factorization in the $[12]\toZ$ channel requires,
\begin{eqnarray}  
\lim_{[12]\toZ} \ap{12}[12]A_{BCF}(1^-,2,3,4) &=& \lim_{[12]\toZ} \ap{14}^2\ap{13}^2[34]^2 \gamma_{234} ( \frac{g_{44}-g_{33}}{su} -\frac{g_{33}}{tu} ), \nonumber \\
&=& \ap{14}^2\ap{13}^2[34]^2 \gamma_{234}g_{22}\frac{1}{t^2} .
\end{eqnarray}
Again, there is a potential double pole. The double piece vanishes and
the correct factorization limit is obtained provided
$g_{22}=g_{33}=g_{44}$. This is a special case of the principle of equivalence
for gravity --interacting matter couples to gravity with the same
strength. 

\section{Factorization of Higher-Point Amplitudes}\label{sec:4}
In this section, we will show that $n$-point amplitudes obtained by
BCFW recursion satisfy the physical requirements of complex-momentum unitarity (factorization).  We have already noted
that Lorentz invariance is manifest in the construction, so long as
the recursion begins from Lorentz-invariant primitive three-point
amplitudes.  We now describe the factorization requirement in more
detail, for a theory of massless particles at tree-level.

In the limit that the summed momentum of a collection of legs, $P_I =
\sum_{i \in I} p_i$ becomes null, unitarity requires that
the amplitude have a single pole, whose residue is a product of two
sub-amplitudes: \be \lim_{P_I^2 \toZ} P_I^2 A(1,\dots, n) =
\sum_{h,a}A(I, -P_I^{-h,a}) A(P_I^{h,a}, \bar I),
\label{eqn:factorization}
\ee 
where $\bar I$ is the set of legs \emph{not}
in $I$, which must contain at least two legs (we can label the same
limit equivalently by either $I$ or $\bar I$).  When $I$ contains only
two legs, \be (p_i + p_j)^2 = 2p_i \cdot p_j = \lambda_i
\lambda_j\,\tilde \lambda_i \tilde \lambda_j = \ap{ij} [ij] \ee goes
to zero if either $\ap{ij}\toZ$ or $[ij]\toZ$; for complex momenta we
can obtain limits in which one spinor product goes to zero while
the other remains finite.  This results in distinct factorization
formulas, involving $A^{(a)}(i,j,-P_{ij})$ when $\ap{ij}\toZ$ and
$A^{(h)}(i,j,-P_{ij})$ when $[ij]\toZ$.  

\bef
\includegraphics[height=0.9in]{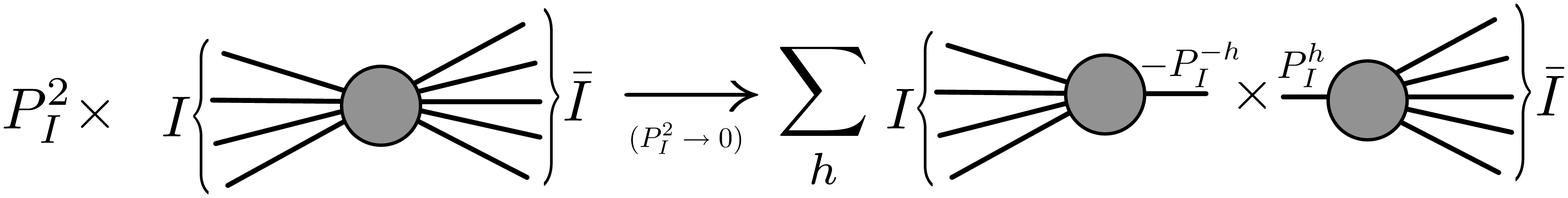}
\caption{A diagrammatic picture for the requirement of complex factorization.\label{fig:ComplexFactorization}}
\eef

\bef
\includegraphics[width=6.5in]{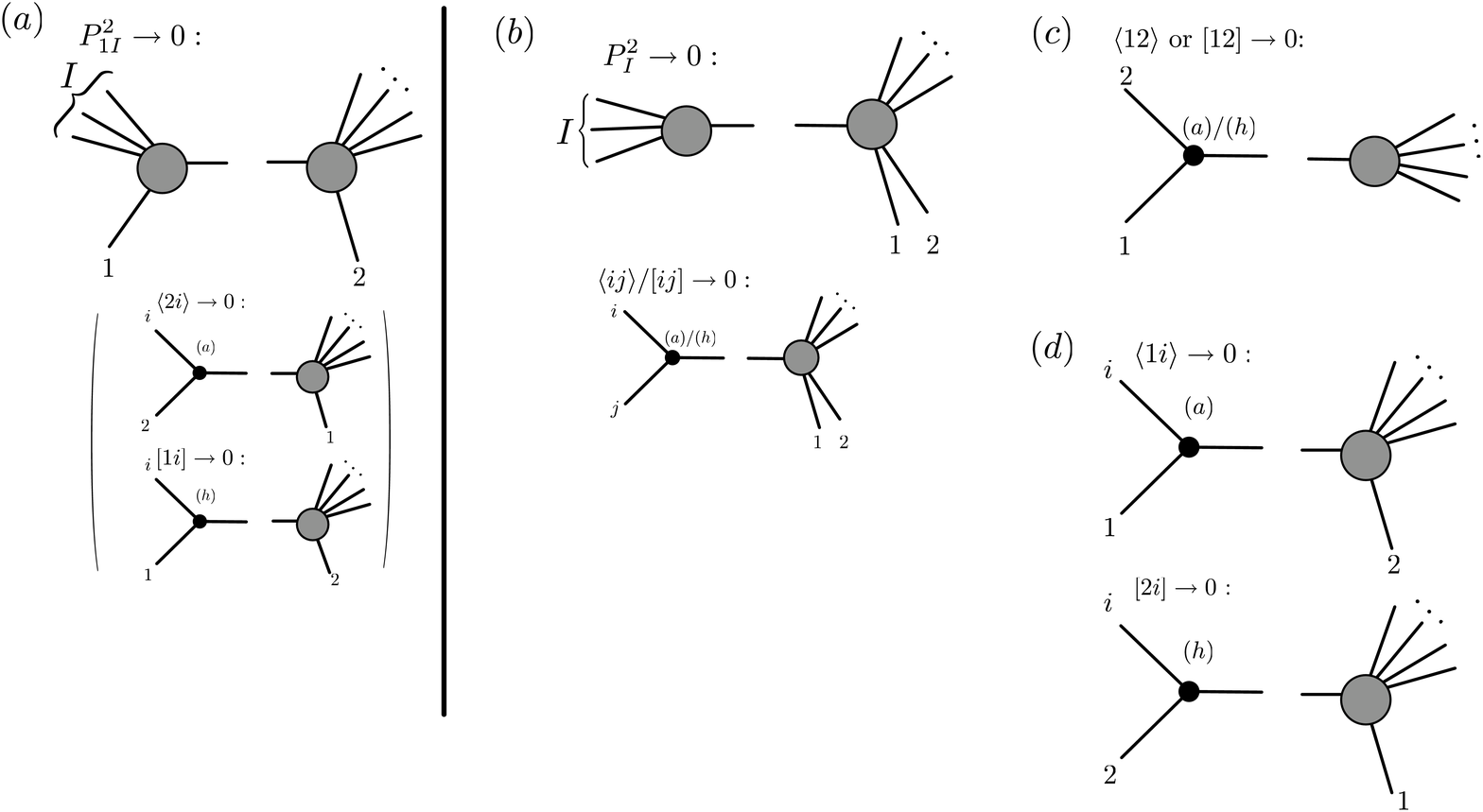}
\caption{A graphical classification of poles in a general amplitude.
  The BCFW construction manifestly has proper 
  factorization on the poles $(a)$ to the left of the
  thick line.  The three classes of diagrams to the right of the line
  must be checked explicitly.  We discuss the ``unshifted'' poles
  $(b)$ in Sec.~\ref{sec:bigunshifted} (the small diagram below is a
  special case).  The remaining two-particle poles correspond to
  invariants that involve one or more
  BCF-shifted legs, but nonetheless are not altered by BCF shifts.
  The ``unique diagram'' poles shown in $(c)$ are discussed in
  Sec.~\ref{sec:12pole}, and the ``wrong-helicity'' poles $(d)$ in
  Sec.~\ref{sec:wrongfac}.  Notable features of each pole are
  discussed in the text. \label{fig:BCFsummary}}
\eef

The BCFW recursion singles out the two shifted legs, which we will
label 1 and 2, and treats 
different poles differently.  The BCFW formula \eqref{eqn:bcf12}
manifestly satisfies the factorization requirement
\eqref{eqn:factorization} in each of the poles that appears in the
sum, namely $P_{1I}^2 \toZ$ for some set of legs $I$.  The explicit
propagator $1/K^2$ in the term of \eqref{eqn:bcf12} with $L=I$ becomes
singular.  Nothing else in the BCFW expression can be singular in this
limit --- the sub-amplitudes are evaluated at shifted momenta, at
which no kinematic invariants inside the sub-amplitudes vanish.  If
either side of the pole has only two elements, we must specify which
spinor product is exposed by the BCFW shift: if $I$ has only one
element $i$, the BCFW shift exposes the pole $[1i]\toZ$, and if $\bar
I$ contains only the legs ${2,j}$, the BCFW shift exposes the pole
$\ap{2j}\toZ$.

Factorization in the remaining poles, which correspond to kinematic
invariants that do not depend on $z$, must be verified explicitly.
Checking that BCFW amplitudes factorize in these non-manifest limits
is the main task of this section; we will demonstrate it by induction,
assuming that $(n-1)$-point amplitudes factorize.  The arguments are
somewhat technical, but contain surprising structure; we will
highlight the pieces of the BCFW expression that \emph{do} contribute
to each singularities, and the structural properties of the amplitudes
that are required for factorization.  We now classify these poles, and
summarize the properties required for their factorization.

We first consider poles $P_I^2 \toZ$ where $1,2 \in \bar I$ and $\bar
I$ contains at least one additional leg.  Terms in the BCF sum for
which $I\subset L$ or $I \subset R$ are singular in this limit;
because the BCFW shift does not affect the kinematics of the legs in
$I$ in any way, the inductive proof of Section \ref{sec:bigunshifted}
is straightforward, and independent of all details of the chosen
amplitudes and BCFW shift.  When $I$ contains only two legs, one
subtletly will appear that requires the same relations between coupling
constants seen in the four-point amplitudes of Section \ref{sec:consistency}. 

BCFW shifts leave two other types of invariant unaffected: $[12]$
(and $\ap{12}$) and $\ap{1i}$ ($[2i]$) for $i \neq 1,2$.  Unlike those
discussed above, factorization in these limits depends on the spins
and helicities of the particles involved.  The collinear singularity
as $[12] \toZ$ arises in the BCFW formula from amplitudes where a
\emph{soft} BCFW-shifted leg 1 is attached to each of the un-shifted
legs (Sec.~\ref{sec:12pole}).  This limit is closely related to the
soft-photon and soft-graviton limits considered by Weinberg
\cite{Weinberg:1964ew}.  For spin-1, color-ordering reduces the factorization
statement to the equality of one soft and one collinear diagram or of
two soft diagrams.  The spin-2 factorization receives contributions
from soft singularities in $(n-2)$ terms, each of which produces a
\emph{double} pole in $\ap{12}$; the correct single pole of
factorization is obtained using momentum conservation.  Attempting to
apply the BCFW recursion shifting legs with helicities $\sh{+s,-s}$,
which is not a valid BCFW shift, we see that factorization cannot be
satisfied.  

The ``wrong-helicity'' factorization limit $\ap{1i}\toZ$, which we
discuss in Section \ref{sec:wrongfac}, arises from multiple BCFW
terms in both gauge theory and gravity.  In gauge theory, their sum
can be interpreted precisely as a BCFW construction of an
$(n-1)$-point amplitude.  However, the helicities of the shifted legs
in the $(n-1)$-point amplitude can differ from those in the original
$n$-point amplitude, so for example the proof of factorization for
$\sh{++}$ shifts depends on the validity of $\sh{-+}$ shifts as well.
The proof also depends on the large-$z$ scaling properties of
amplitudes, in particular the vanishing of amplitudes as $1/z^s$ at
large $z$, and the growth of amplitudes under ``invalid'' shifts
$\sh{+-}$ bounded by $z^{3s}$.  A simple power-counting argument justifies
this scaling for $s=1$ (see Appendix \ref{app:scaling}), but
\textbf{we do not have a proof of this scaling for gravity, so the
  factorization argument in that case remains incomplete.}  As for the
$\ap{12}$ poles, factorization is violated for the invalid shifts, in
this case by non-vanishing boundary terms.

\subsection{Factorization on Unshifted Multi-Leg Poles $P_I^2
  \toZ$}\label{sec:bigunshifted}
Let $I$ be a set of legs that excludes the shifted legs 1 and 2, and at
least one other leg. 
Factorization requires that in the limit
$P_I^2 \rightarrow 0$,
\be 
P_I^2 A(1,2,\dots,N) \rightarrow \sum_{h_I} A(\bar I, P_I^{h_I})
A(-P_I^{-h_I}, I),
\label{unshifted-pole-ans}
\ee
where $\bar I$ is the complement of $I$ in ${1,2,3,4,\dots n}$. 

If $I$ contains three or more legs, this result is readily obtained by
considering the limit as $P_{I}^2 \toZ$ of the BCFW
decomposition \eqref{eqn:bcf12} of $A(1,2, \dots,n)$ (one subtlety
will arise in the two-particle case, discussed below).  The only terms
that are singular in these limits are those with  $I \subset L$ or $I
\subset R$.  Factorization of
the left or right 
\emph{sub-amplitudes} (the induction hypothesis) yields a simple limit:
\be K_I^2 A(\hat 1, I, \tilde L, \hat K)
\rightarrow \sum_{h_I} A(\hat 1, \tilde L, \hat K, P_I^{h_I}) A( - P_I^{-h_I}, I),
\label{subfactor}
\ee
where $\tilde L$ is the subset of $L$ not explicitly written; terms
with $I$ in the right factor have analogous limiting behavior.  In particular,
all of these terms have a common factor $A(\hat P_I,I)$, which is one
of the factors of the desired $n$-point factorization limit.
Adding these terms, we find
\begin{align}
P_I^2 A_{BCF} & \rightarrow 
\sum_{h_{K}, h_I} & A(I, -\hat P_I^{-h_I})  \sum_{L/R} \bigg( 
A(\hat 1, \tilde L, P_I, \hat K)  \f{1}{K^2} A(-\hat K, \hat 2,
\tilde R)\qquad\qquad
 \label{line1}\\
& & + A(\hat 1, \tilde L, \hat K)  \f{1}{K^2}  A(-\hat K, \hat 2,
 \tilde R, P_I) \bigg) \qquad \label{line2}\\
&=  \sum_{h_I} & A(P_I{-h_I}, I) A(\bar I, P_I^{h_I}),\hspace{2.3in}
\label{longsum}
\end{align}
where in the last line, we have recognized terms \eqref{line1} and
\eqref{line2} as a BCF formula for the lower-point amplitude 
$A(\bar I, P_I)$. This
argument is illustrated diagramatically in Figure \ref{fig:bigunshiftedDiagram}.

\bef
\includegraphics[width=5in]{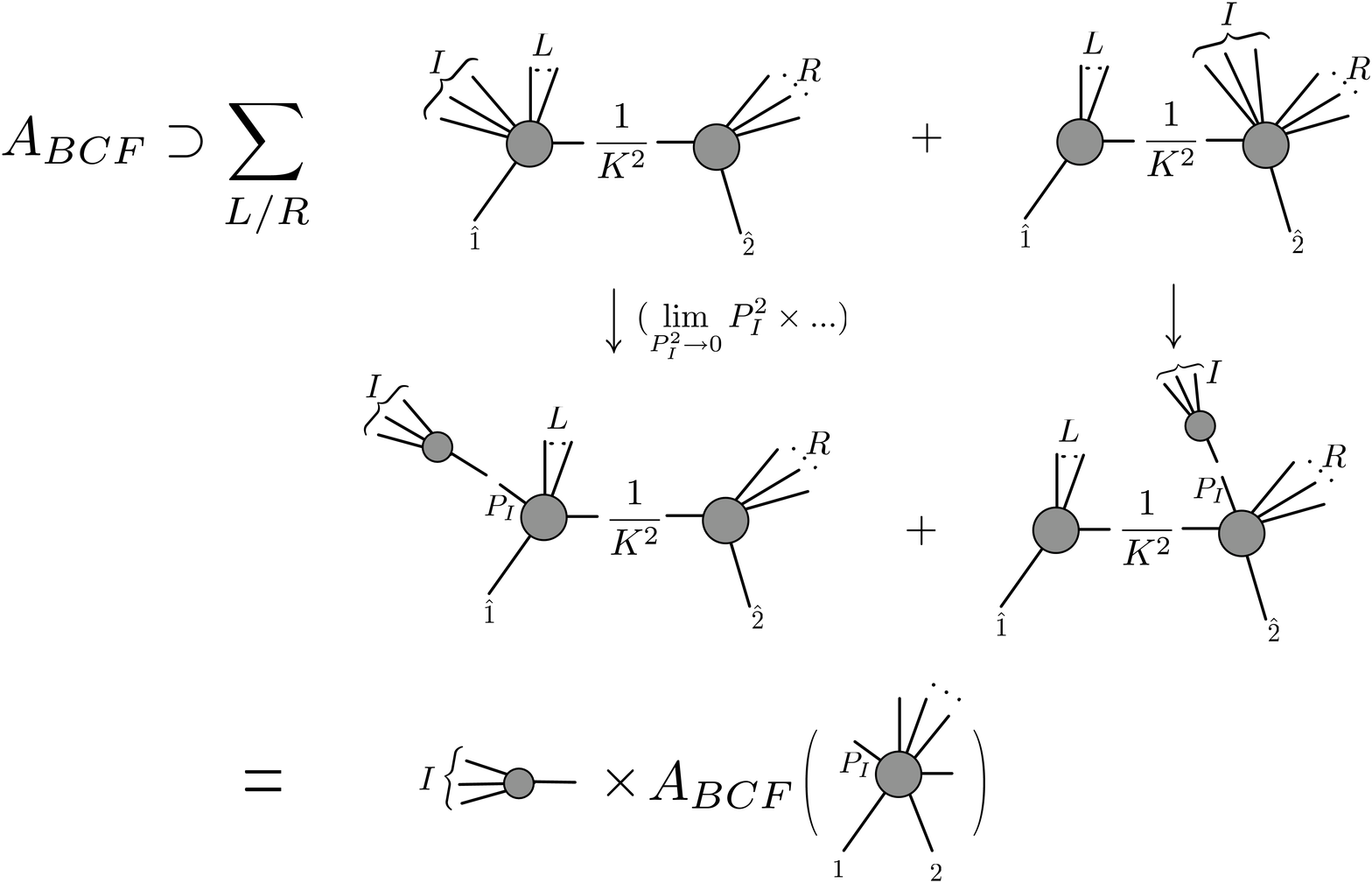}
\caption{A graphical summary of the argument for factorization of the
  BCFW amplitudes in unshifted poles.  The top line identifies the
  subset of terms in the BCFW recursion relation that are singular as
  $P_I^2 \toZ$ (those in which all legs of $I$ are on the same side).
  Their singularities are shown on the second line; pulling out a
  common factor $A(-P_I^{-h_I}, I)$, we recognize the sum as the
  BCFW-recursed expression for the second factor in
  \eqref{unshifted-pole-ans}. \label{fig:bigunshiftedDiagram}} \eef

Note that the argument above fails completely if $\bar I$ contains
only one and two --- there are no diagrams in the BCFW sum for which
$I \subset L$ or $R$, and in fact, as we will see, very different
terms are singular in that case, which we consider in
Section \ref{sec:12pole}.  We first elaborate on the special
case that the set $I$ contains only two particles ($i,j$); in this
case, one additional class of diagrams plays a role.

\subsubsection{Two-Particle Unshifted Poles}
For definiteness, we consider the singularity $[34]\toZ$.  In this
case, one term in the BCFW expansion that we might expect to be
singular is in fact non-singular, and two new terms are singular:
\begin{itemize}
\item One of the terms in the $n$-point BCFW sum \eqref{longsum} is
\be
A(\hat 1, 3, 4, \hat K) \frac{1}{K_{134}^2}  A(-\hat K, \dots, \hat
2).
\label{nosing}
\ee
The singularity of the four-point sub-amplitude as $[34]\toZ$ is 
$A^{(h)}(3,4,-P_{34}^{-h_{34}}) \times A^{(a)}(\hat 1,
P_{34}^{h_{34}}, \hat K)$, but as $[34]\toZ$,  $[\hat 1
  K_{34}]$ also approaches zero.  Therefore the $A^{(a)}$ factor vanishes for
both gauge theory and gravity, and \eqref{nosing} does not contribute
to the singularity in the factorization limit $[34]\toZ$.

\item Two additional terms in the $n$-point amplitude, not included in \eqref{longsum}, are singular ---
 those in which the left factor contains only two legs: $\hat 1$ and
 either 3 or 4.  When legs $\hat 1$ and $3$ are alone in
the left factor, for instance, the intermediate leg $K$ has $|\hat K] \propto
|3]$, so that as $[34]\toZ$ we also have $[\hat K 3]\toZ$; therefore,
this term can be singular even though legs $3$ and $4$ are split
between factors.  The singularity of this term is given by
\be
\sum_{h_{13},h^*} 
A^{(h)}(\hat 1, 3, -\hat
    K_{13}^{-h_{13}}) \f{1}{K_{13}^2} 
\left\{ \f{\ap{34}}{\ap{\hat K_{13} 4}} A(\hat K_{13}^{h_{13}}, 4 ,
P_{\hat 1 3 4}^{h^*}) A(P_{\hat 1 3 4}^{-h^*}, \dots,
    \hat 2), \right\}
\label{squareIJ_newTerm}
\ee
where the factor in braces is obtained by factorization of the right
BCFW factor.  A similar term is obtained when legs 3 and 4 are exchanged.
\end{itemize}

We recall the argument for factorization in multi-particle poles:
the factorization limits of every individual term in the $n$-point BCF
amplitude can be interpreted as a contribution to the BCFW formula for
the lower-point ampitude $A(P_I, \bar I)$ appearing in
\eqref{unshifted-pole-ans}.  By the discussion above, there are two
new terms in the BCFW ansatz that do not have this form.  There is
also a term that must appear in the BCFW formula
(\ref{line1}-\ref{line2}) for the lower-point amplitude, but is not
generated by factorizing any one term in the $n$-point BCFW ansatz.
This missing term is
\be
    \sum_{h_{34}, h_K}A^{(h)}(3, 4,
    -P_{34}^{-h_{34}}) A^{(h)}(\hat 1, P_{34}^{h_{34}}, \hat
    K^{h_K}) \frac{1}{K_{134}^2} A(- \hat K^{-h_K}, \dots, \hat 2)\label{rightSquareIJ} 
\ee
(we maintain the conventions introduced earlier, that $P$'s denote
summed momenta that become null in singularity limits, $K$'s unshifted
intermediate momenta in the BCFW formula, and $\hat K$ the 
null shifted intermediate momenta in BCFW).

To recover the correct factorization result, the extra terms in the
factorization of the $n$-point BCFW formula must compensate for the
missing term --- Eqn. \eqref{squareIJ_newTerm} and its analogue with 3
and 4 exchanged must sum to \eqref{rightSquareIJ}.  This is so, and
follows from the Jacobi identity (spin-1) or equality of all couplings
(spin-2).  It is worth stressing that, although our result will be
very reminiscent of the factorization of four-particle amplitudes in
Section \ref{sec:consistency}, it is not the same physical limit ---
indeed, the products $A^{(h)} A^{(h)}$ do not arise as factorization
limits of four-point amplitudes.  In fact, these products are non-zero
only for external helicities $(-,-,-,+)$, for which the four-point
amplitude vanishes.

To proceed, let us define 
\be
f(i,j;k,l) = \sum_{h_{int}} A^{(h)}(i,j,K_{int}) \times
A^{(h)}(-K_{int}, k, l) 
\ee 
The $f$'s have a simple form: 
\be
f(i,j;k,l) = \left(\sum_a f_{ija}f_{akl}\right)
\left(\frac{1}{\ap{1j}\ap{kl}} H(i,j,k,l)\right)^s, 
\ee 
where 
\be
H(1^-,2^-,3^-,4^+) = \frac{[4\mu]^3}{[1\mu][2\mu][3\mu]} 
\ee
depends only on the helicities of the particles (not on how they are paired),
and is zero for all other combinations.

Terms \eqref{squareIJ_newTerm} and \eqref{rightSquareIJ} all have
the limiting form
\be
\sum_{h_{int},h_P} \mbox{(kinematics)}\times f(i,j;k,P^{h_P}) \times A^{(n-2)}(-P^{-h_P}, \hat
2, \dots),
\ee
where ${i,j,k}$ is some ordering of ${\hat 1,3,4}$, where $\hat 1$ is
the limiting BCF-shifted momentum of leg $1$ in \eqref{rightSquareIJ},
and $P = K_{34} + \hat p_1$ (all three terms approach this uniform
kinematics as $[34]\toZ$).  Factorization requires the coefficients of
$A^{(n-2)}$ in \eqref{squareIJ_newTerm} $(+\; 3 \leftrightarrow 4)$ to
reproduce the coefficients in \eqref{rightSquareIJ}, i.e.
\be
-\frac{1}{K_{134}^2} f(3,4;\hat 1,P) + 
\frac{1}{K_{13}^2}  \f{\ap{34}}{\ap{\hat K_{13} 4}} f(\hat 1,3;4,P) +
\frac{1}{K_{14}^2}  \f{\ap{43}}{\ap{\hat K_{14} 3}} f(\hat 1,4;3,P).
\ee
By judicious use of kinematic identities, this can be rewritten as
\be
\frac{\ap{34}}{[2P]} H(1,3,4,P)^s
\left\{
(\ap{1P}{34})^{s-1} f_{1Pa}f_{a34}+
(\ap{13}{4P})^{s-1} f_{13a}f_{a4P}+
(\ap{14}{P3})^{s-1} f_{14a}f_{aP3}
\right\},
\ee
which vanishes by the Jacobi identity for $s=1$ and the Schouten
identity (setting all $f$ equal) for $s=2$.

Thus, the non-standard terms \eqref{squareIJ_newTerm} are equal 
to the term \eqref{rightSquareIJ} that was missing from the naive sum,
and factorization on poles like $[34]\toZ$ is guaranteed by the BCFW
construction for both spin-1 and spin-2.  An analogous result would
hold when the opposite-helicity invariants $\ap{ij}\toZ$, except that
in that case, the roles of legs 1 and 2 are interchanged, and the
identity involves anti-holomorphic 3-point amplitudes instead of
holomorphic ones.

\subsection{Factorization on ``Unique Diagram'' Poles
  ($[12]$)}\label{sec:12pole} 

The BCFW formula does not contain any
sub-amplitudes with ``propagator'' singularities as $[12] \toZ$
because, by construction, legs 1 and 2 are in separate factors and the
remaining $(n-2)$ legs are split between the factors.  Instead, the
collinear singularities $[12]\toZ$ arises remarkably in the BCFW
formula through a \emph{soft} singularity in one of the factors.
Specifically, whenever the left-hand factor in the BCFW decomposition \eqref{eqn:bcf12} is a three-point
amplitude $A^{(h)}(\hat 1, i, -\hat K)$ for some leg $i$, we find $|\hat 1]
= \fsp{21}{2i}|i] \rightarrow 0$ (analogously,
in the $\ap{12}\toZ$ limit, $\rr{\hat 2}$ becomes soft when the
right-hand factor is a three-point amplitude).  The right factor, an
$(n-1)$-point amplitude, approaches a uniform kinematic limit for all
$i$, which is in fact exactly the kinematics that appears in the
factorization formula:
\be
\lim_{[12]\toZ} \ap{12}[12] A(1,\dots,n)
= \sum_{h'}A^{(h)}(1,2,-P_{12}^{-h'}) 
A(P_{12}^{h'},3,\dots,n).
\label{eqn:12fact}
\ee

The behavior of the three-point amplitude $A^{(h)}(\hat 1, i, -\hat
K)$ as $|\hat 1]\toZ$ depends on the helicity $h_1$ of the soft leg.
If $h_1=+s$, all three-point amplitudes $A^{(h)}$ in
\eqref{eqn:12fact} vanish as $|\hat 1]\toZ$, so the BCFW sum is
non-singular.  This makes the factorization limit either trivial (when $h_2=+1$, the three-point amplitude in \eqref{eqn:12fact} is also zero) or
manifestly violated (when $h_2=-1$, \eqref{eqn:12fact} is non-zero,
but the BCFW formula cannot reproduce this singularity).
The latter case corresponds to the shift $\sh{+,-}$, which we have
already seen does not produce consistent four-point amplitudes.

\bef
\includegraphics[width=6in]{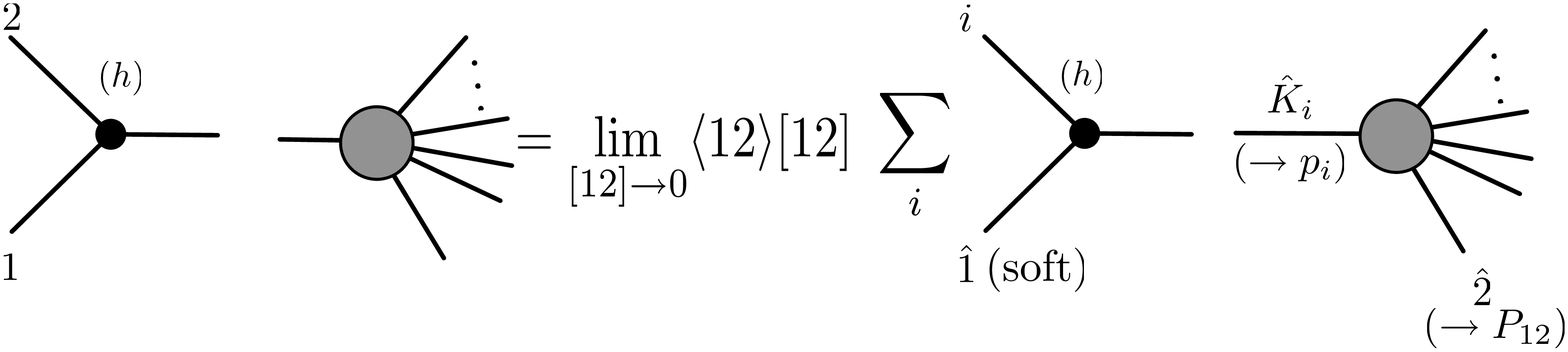}
\caption{Left: factorization limit for the  ``unique diagram'' pole
  $[12]\toZ$.  Right: in each of the BCFW terms that contribute to this
  singularity, $\hat 1$ is a \emph{soft} line attached to one of the
  unshifted legs. \label{fig:UniqueDiagramPoles}}
\eef

If $h_1=-s$, the three-point amplitudes in \eqref{eqn:12fact} do have
a soft singularity (they scale as $[12]^{-s}$), and the factorization condition is non-trivial.
In this case, the singularity of the BCFW amplitude is given by
\bea
\lim_{[12]\toZ} \ap{12}[12] A_{BCF}(1,\dots,n) = 
\lim \sum_{i,h_{1i}} & \left( \frac{\ap{12}[12]}{\ap{1i}[1i]} 
A(\hat 1^{-s,a_1},i^{h_i,a_i},-\hat K_{1i}^{-h_{1i},b}) \right)
\nonumber \\
& \times A(\hat
2,\dots,i-1, \hat K_{1i}^{h_{1i},b},i+1,\dots,n),
\label{eqn:iterm}
\eea
where we have written helicity and species indices explicitly in the
three-point amplitude and for the modified leg $K_i$ in the
$(n-1)$-point amplitude. 

We begin by describing the kinematics of \eqref{eqn:iterm} more
explicitly to illustrate the soft limit, then consider the sums
separately for the spin-1 and spin-2 cases.
The  shifted
spinor $|\hat 1] = |1] + z_* |2]$ in the $i$'th term is proportional
to $|i]$;  the
constant of proportionality and the value of $z_*$ can be found by
taking the inner product of $|1] + z_* |2] =c |i]$ with $|2]$ or
$|i]$, respectively:
\be
|\hat 1] = \fsp{12}{i2} |i] \quad \mbox{ at } z_* = - \fsp{1i}{2i}.
\ee
We note that $|\hat 1]$ is indeed becoming soft (one can
understand the limit as follows: as $|1]$ and $|2]$ become nearly
proportional, a fine-tuned subtraction of nearly equal spinors is required to
obtain a spinor $|\hat 1]\propto |i]$).  Since $|\hat 1]$ is soft, the
momentum $\hat K_i$ leaving the $(n-1)$-point factor must approaches
$p_i$:
\be
\hat K_i = (\rr{i} + \fsp{12}{i2} \rr{1}) |i] ,
\ee
and the shifted  momentum of leg 2 is,
\be
\left(\rr{2} + \fsp{1i}{2i}\rr{1}\right)\,|2] \rightarrow P_{12} = (p_1+p_2).
\ee 
As we have noted, the limiting kinematics of the $(n-1)$-point
amplitudes as $[12]\toZ$ is independent of which leg $i$ appears in
the left factor!  In fact, it is the same kinematics
that appears in  the $(n-1)$-point amplitude of \eqref{eqn:12fact}.
However, the $\hat K_{1i}$ line leaving the $(n-1)$=point diagram in
each case can have different internal species quantum numbers than those of the
original outgoing line $i$.  

One might also expect the line $K_i$ to have different helicity than
$h_i$, but in fact these diagrams need never be considered.  We have
already noted that, if $h_1 = +s$, the factorization requirement is
either trivial (when $h_2=+s$) or impossible to satisfy ($h_2=+-$);
the non-trivial case is $h_1 -s$.  But the gauge/gravity amplitude
$A^{(h)}(-s, -s, +s)$ is only non-zero if its two remaining legs have
opposite helicities, which in our notation is $h(K_i)=h_i$.  

Before considering the singular three-point terms in more detail, we
clarify why the generic case --- in which the left factor is a
higher-point amplitude, does not contribute.  In this case, as
$[12]\toZ$, the shifted leg $|1]$ approaches a non-singular limit
\be
|\hat 1] = - \frac{K_J^2}{2 p_1.K_J} |1],
\ee
where $K_J$ is the (generically non-null) sum of momenta of the legs
in the left factor.  As no momenta or invariants within either
sub-amplitude vanish at this point, we expect no singularities.
Likewise, no singularity occurs when the right-hand factor is
three-point, because  $\rr{2}$ and the spinor $\rr{1}$ by which it is
shifted are not orthogonal.
 
\subsubsection{Spin-1 Factorization}
We now consider the three-point amplitudes in somewhat more detail for
the theories of pure spin-1 ``gauge-theory-like'' (++- and --+)
interactions.  The kinematic factor times three-point amplitude in the $i$'th term
(in parentheses in Eqn. \eqref{eqn:iterm}) is given by   
\be
%  \frac{\ap{12}[12]}{\ap{1i}[1i]} 
% A^{(h)}(\hat 1^{-s,a_1},i^{h_i,a_i},-\hat K_i^{-h_i,b}) = 
 \frac{\ap{12}[12]}{\ap{1i}[1i]} 
\left(\ap{1i}  \fsp{i2}{12}\right) f_{a_1,a_i,b}=
-\ap{12} \fsp{2i}{1i} f_{a_1,a_i,b}.\label{eqn:12gaugeLimit}
\ee
Note that $\fsp{2i}{1i}$ approaches an $i$-independent, finite limit
as $[12]\toZ$, so different terms in the sum \eqref{eqn:iterm} differ
only in the replacement of the $i$'th particle's species index by a
dummy label $b$, and contraction into a coefficient
$f_{a_1,a_i,b}$.  One can see that this identity holds by
separately considering every contraction of many $f$'s that could
appear in the $(n-1)$-point amplitudes in \eqref{eqn:iterm}, and
repeatedly using the Jacobi identity.  

It is much less cumbersome, however, to switch to the color-ordered formalism
\cite{Berends:1987cv,Mangano:1987xk,Mangano:1988kk}.  Because the cyclic ordering of leg indices is
significant in color-ordered amplitudes, we will call the BCFW-shifted
legs $\sh{x,y}$ in this discussion rather than $\sh{1,2}$, to avoid
suggesting that they are color-adjacent when they need not be.  We
have already seen that, for four-point spin-1 amplitudes to factorize,
three-point couplings $f$ must satisfy a Jacobi identity.  We can
then associate each ``species index'' $a$ with an element $T_a$ in the
adjoint representation of a Lie algebra; the relation on species
indices described above is a trivial consequence of this structure.
The only color structure ever generated in tree amplitudes is a single
trace, allowing us to express any amplitude as a sum of color traces
times colorless primitive (or color-ordered) amplitudes, 
\be
A(1^{a_1}, 2^{a_2},\dots, n^{a_n}) = \sum_{P(2,\dots,n)} Tr(T^{a_1}
  T^{a_{i_2}}\dots T^{a_{i_n}}) \times A_{c.o.}(i_1,i_2,\dots,i_n).  
  \ee
  The color-ordered amplitude receives contributions only from
  diagrams in which the legs are cyclically ordered from 1 to $n$
  clockwise on a plane, and all coefficients $f_{a,b,c}$ can be
  replaced with a uniform coupling constant $f$.  Aside from this,
  color-ordered BCFW amplitudes have the same structure as unordered
  ones.  The color-ordering is, for our purpose, simply a bookkeeping
  device to focus on one color-structure (i.e. one set of contracted 
  $f$'s) at a time using their known algebraic structure.  This
  bookkeeping simplifies the discussion considerably: instead of $n$
  diagrams, for a given color-ordering only two diagrams contribute to
  this discussion.

We must consider two cases separately: when the legs $x$ and $y$
are color-adjacent, and when they are not.  If legs $x$ and $y$ are
not color-adjacent, then no diagrams in which $x$ and $y$ connect at a
three-point vertex have the correct color ordering, so there is no singularity as $[xy]\toZ$; there are, however, two
non-vanishing terms in the BCFW expression \eqref{eqn:iterm} (namely, 
$i=x\pm 1$).  These appear with opposite signs, since the
color-ordering includes $A(x-1, x, -K) = - A(x,x-1,-K)$ and
$A(x,x+1,-K)$ (with no sign flip).  Thus the two diagrams cancel, correctly giving no singularity.

If $x$ and $y$ are color-adjacent (for definiteness, say $x=1$,
$y=2$), the factorization limit \eqref{eqn:12fact} is non-zero:
\be
A^{(h)}(1^{-1},2^{h_2},-P^{-h_2}_{12}) A(P_{12}^{h_2},3,\dots,n) = \ap{12} \fsp{2\mu}{1\mu}
 A(K_{12},3,\dots,n),
\ee
and is reproduced exactly by the one non-zero term in the
color-ordered BCFW expression (using \eqref{eqn:iterm} and antisymmetry
of the three-point amplitude),
\be
 \frac{\ap{12}[12]}{\ap{1i}[1i]} 
A^{(h)}(n, \hat 1,-\hat K_n) 
A(P_{12},3,\dots,\hat K_n).
\ee

In each case, then, the color structure established at four-point and
properties of three-point amplitudes suffice to guarantee the
factorization of $n$-point BCFW amplitudes as $[12]\toZ$ (the
analysis for $\ap{12}\toZ$ is analogous, with the roles of legs 1 and
2 exchanged).

\subsubsection{Spin-2 Factorization}\label{spin2_12}
To study spin-2 interactions, we use the fact derived from spin-2
four-point amplitudes in \cite{Benincasa:2007xk} that interactions
among spin-2 particles can always be written as self-interactions of a
single species. Therefore, there are no species indices in
\eqref{eqn:iterm}, and the $(n-1)$-point amplitudes approach truly
identical limits.  Unlike the case of spin-1, however, the kinematic
$\times$ three-point factors of \eqref{eqn:iterm} do depend on the $i$'th
particle's momentum --- they are given by, 
\be
% \frac{\ap{12}[12]}{\ap{1i}[1i]} 
%A^{(h)}(\hat 1^{-s},i^{h_i},-\hat K_i^{-h_i}) = 
 \frac{\ap{12}[12]}{\ap{1i}[1i]} 
\left(\ap{1i}  \fsp{i2}{12}\right)^2=
\frac{1}{[12]}
 \frac{\ap{12}[2i]^2\ap{1i}}{[1i]}.
\ee
Moreover, the individual terms become singular as $[12]\toZ$
(i.e., since we are attempting to evaluate a residue, each BCFW term
in fact has a \emph{double} soft singularity as $[12]\toZ$).
Replacing each of the $(n-1)$-point amplitudes in \eqref{eqn:iterm}
with its limiting kinematics, we
find, 
\be
\lim_{[12]\toZ} \ap{12}[12] A_{BCF}  =A^{(n-1)} \times \lim_{[12]\toZ}
 \left(\sum_{i}
\frac{1}{[12]}
 \frac{\ap{12}[2i]^2\ap{1i}}{[1i]}\right).\label{eqn:gravlimit}
\ee
Writing, 
\be
\frac{[2i]^2}{[1i]} = \fsp{2\mu}{1\mu} \left( [2i] +
\fsp{12}{1\mu} [\mu i] \right) + O([12]^2),
\ee 
the limit of \eqref{eqn:gravlimit} becomes,
\be
\fsp{2\mu}{1\mu} \left[
\lim \frac{\ap{12}}{[12]} 
     \left(\sum_{i}[2i]\ap{1i}\right)
+  \frac{\ap{12}}{[1\mu]} \left( \sum_{i} \ap{1i}[\mu i] \right)\right].
\ee
By momentum conservation, the first sum in parentheses is identically
zero and the second is equal to $\ap{12}[2\mu]$; thus we recover the
singularity limit,
\be
A^{(n-1)} \times \ap{12}^2 \left(\fsp{2\mu}{1\mu}\right)^2.
\ee

In fact the argument above is a bit too quick: for small but finite $[12]$,
each $(n-1)$-point amplitude in \eqref{eqn:iterm} is evaluated at
slightly different kinematics; because the prefactors
in \eqref{eqn:iterm} are themselves growing as $\frac{1}{[12]}$, this
displacement could change the final result by a non-zero amount.  As
we show explicitly in Appendix \ref{app:derivatives}, this correction
has no effect at all on the limit --it is proportional to a sum over
all legs $i$ of $[\mu i] [\mu| \frac{dA^{(n-1)}}{d|i]}$, which vanishes
because $|i] dA/d|i]$ is antisymmetric.

\subsection{Factorization on Wrong-Factor Poles ($\ap{1j}$)}
\label{sec:wrongfac}
The final class of factorization limits we must check is the limit
$\ap{1j} \toZ$ for $j \neq 2$ (the case $[2j]\toZ$ for $j\neq 1$ is
analogous).   We will consider concretely the limit $\ap{13}\toZ$
(again, the sequence of labels is arbitrary).
In this limit, we require
\be
\lim_{\ap{13}\toZ} [1 3] \ap{13} A(1,3,\dots,n)
= \sum_{h_{13}} A^{(a)}(1, 3, -P_{13}^{-h_{13}}) A(P_{13}^{h_{13}},2,\dots,n).
\label{eqn:13fact}
\ee
As before, we will prove this for $n$-point amplitudes by induction,
assuming that all lower-point amplitudes factorize appropriately and
can be generated by the BCFW construction.  

We will classify terms of the BCFW sum as in Figure
\ref{fig:WrongFactorPoles} (we will mention only terms that have
potential singularities).
We will first consider
terms (a), in which the
left factor contains legs 1 and 3, and at least one additional leg ---
these will combine into an expression like \eqref{eqn:13fact}, in
which the $(n-1)$-point amplitude has been expressed using BCFW
recursion.   Terms (b) and (c) will not contribute to the 
factorization limit, but to show this we will need to use the fact
proved in Appendix \ref{app:scaling} for spin-1, that at large shifts
$z$, \emph{$n$-point amplitudes generated by BCFW have the same
$z$-scaling as 3-point amplitudes} (i.e. $z^{-1}$ for all ``legal''
BCFW shifts and no faster than $z^3$ for the ``illegal'' shift
$\sh{+,-}$).  

The outline of this proof holds for gravity as well, but the sum of
terms (a) will contain additional contributions, that only vanish if
the $(n-1)$-point amplitudes fall as $z^{-2}$ under valid BCFW
shifts. Likewise, the absence of singularities from terms (c) only
vanish if multi-point gravity amplitudes have the same $z$-scaling as
3-point amplitudes.  We know this to be true of spin-2 BCFW amplitudes
from analysis of the gauge-theory amplitudes \cite{Benincasa:2007qj, ArkaniHamed:2008yf},
but do not have an S-matrix argument for why it must be so.  

\bef
\includegraphics[width=6in]{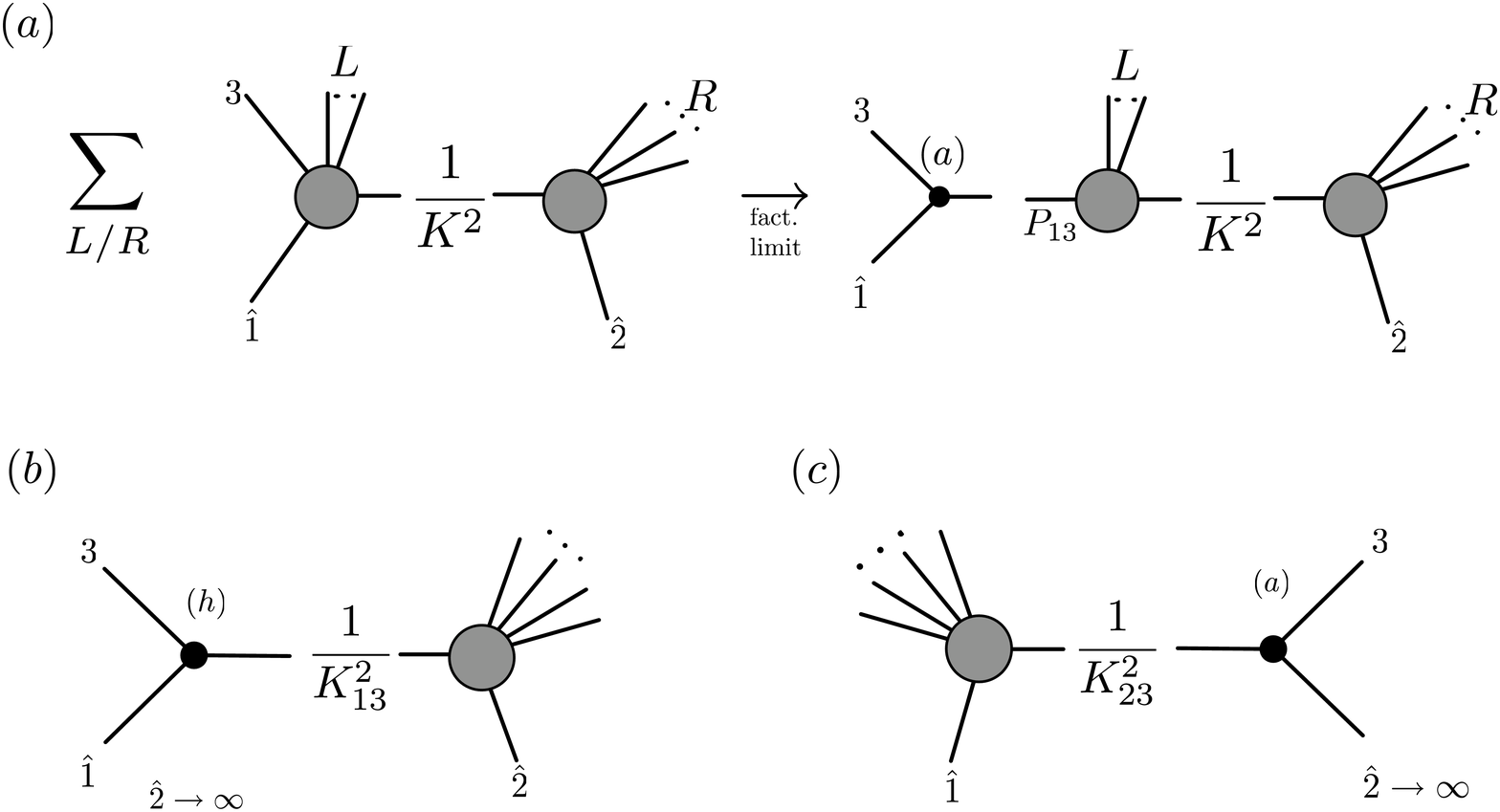}
\caption{Potentially singular terms in the limit $\ap{1i}\toZ$:
Terms of the form (a) reproduce the correct factorization limit.
Terms (b) have no singularity in theories of spin $\ge 1$.  To see
that terms (c) are not singular, one must study the large-$z$ scaling of
lower-point amplitudes.  \label{fig:WrongFactorPoles}}
\eef

\subsubsection{Terms (a)}
Terms of the form (a) in Figure \ref{fig:WrongFactorPoles} are each evaluated at
a different $z=z_*(L)$, but the shift does not change $\rr{1}$ or
$\ap{13}$.  By the induction hypothesis, the left BCFW factor in each
such term should factorize on the $\ap{13}$ pole as, 
\be
\lim_{\ap{13}\toZ} [\hat 1 3] \ap{13} A(\hat 1,3,\tilde L,-\hat K)
= \sum_{h_{13}} A^{(a)}(\hat 1, 3, -P_{\hat 13}^{-h_{13}}) A(P_{\hat
13}^{h_{13}},L,-\hat K).
\label{eqn:13term_b}
\ee
The three-point amplitude in \eqref{eqn:13term_b} vanishes if $h_1 =
h_3 = -s$; otherwise, it is non-zero.  Writing $P_{\hat
13}\rightarrow \rr{1}(|\hat 1] + \frac{\ap{3\mu}}{\ap{1\mu}} |3])$ as
$\ap{13}\toZ$, we find $A^{(a)}\propto [\hat
13]^s \left(\frac{\ap{3\mu}}{\ap{1\mu}}\right)^p$.  The power $p$
depends on the helicities $h_1, h_3, h$ but is the same for all terms;
it is convenient to express all amplitudes in terms of the unhatted
3-point amplitude,
\be
A^{(a)}(\hat 1, 3, -P_{\hat 13}^{-h_{13}}) = \left(\frac{[\hat 1
    3]}{[13]}\right)^s A^{(a)}(1, 3, -P_{13}^{-h_{13}}).
\ee
Substituting this form into \eqref{eqn:13term_b} and summing over all
such contributions to the BCFW formula, we obtain
\bea
\lim_{\ap{13}\toZ} &&[1 3] \ap{13} \sum_{L,h}A(\hat 1,3,
  L,-\hat K_{13L}^{-h}) \frac{1}{K_{13L}^2} A(\hat K_{13L}^h, R, \hat 2)\\
&&= A^{(a)}(1, 3, -P_{13}^{-h_{13}})
 \sum_{L,h} \left(\frac{[\hat 1
    3]}{[13]}\right)^{s-1} A(P_{\hat 13}^{h_{13}},
  L,-\hat K_{13L}^{-h}) \frac{1}{K_{13L}^2} A(\hat K^{h}_{13L}, R,
  \hat 2).
\label{sum13}
\eea
If $s$=1, we recognize the sum in \eqref{sum13} as a BCFW formula for
the $(n-1)$-point amplitude $A(P_{13}^{h_{13}}, 2,4,\dots,n)$ in
\eqref{eqn:13fact}, obtained by a $\sh{P_{13},2}$ shift (note $|\hat
P_{13}] = |P_{13}] + z |2]$ is the same as $P_{\hat 1 3}$ appearing
above).  Thus \eqref{sum13} reproduces the desired factorization limit
\eqref{eqn:13fact}, provided $\sh{P_{13}^{h_{13}},2}$ is a valid shift.

In general, the helicity of $P_{13}$ can differ from $h_1$ (for
instance, if $h_1=+1$ and $h_3=-1$ 
then the only non-vanishing contribution in \eqref{sum13} comes from
$h_{13}=-1$).  So the validity of factorization on $\ap{1j}$ 
poles for shifts $\sh{+,+}$ depends on the validity of a lower-point BCFW formula
using shifts $\sh{-,+}$.  It is important to check that the
above procedure never introduces the BCFW construction using a
$\sh{P_{13}^+,2^-}$ shift.  Indeed, this would require an anti-holomorphic
amplitude $A^{(a)}(1^-, 3^{h}, P_{13}^-)$, which is absent from
the theory.  Thus the shifts valid at 4-point are a closed set.  

Something surprising happens for the spin-2 ($s=2$) case: each term in the would-be BCFW
sum is multiplied by $\frac{[\hat 1 3]}{[13]} = 1 + z_*(L)
\frac{[23]}{[13]}$.  The 1's combine into a BCFW-shifted amplitude,
yielding the correct factorization limit as in the spin-1 case.
However, the sum of terms proportional to $z_*(L)$ remains.  These
are, however, simply the residues of $A(z)$ (recall that the terms in
the BCFW sum are residues of $A(z)/z$, and we have multiplied them by
the $z$'s at which the poles occur)!  therefore, the limit of our
partial sum is given by \be A^{(a)}(1, 3, -P_{13}^{-h_{13}})
\left[A(P_{13}^{h_{13}}, 2,4,\dots,n) + \oint_\infty dz
  A(P_{13}(z)^{h_{13}}, 2(z),4,\dots,n)\right].  \label{eqn:bdry}\ee The contour
integral at infinity vanishes when $A(z)\rightarrow 1/z^2$ --- this is
a known property of gravity amplitudes
\cite{Benincasa:2007qj,ArkaniHamed:2008yf}, but it is surprising from both their
Lagrangian or recursive definitions (we will not prove it here).
The appearance of this formula in the factorization requirement is
mysterious, and suggestive that the $1/z^2$ scaling of gravity
amplitudes is crucial for self-consistency of the theory.

\subsubsection{Terms (b) and (c)}
These are not the only terms in the BCFW sum that can be singular in
the $\ap{13}\toZ$ limit.  We now consider two more potentially
singular terms in the BCFW sum (which we expect to vanish, since we
recovered the correct answer already from terms (a)):
\be
\sum_{h_K}A^{(h)}(\hat 1,3,-\hat K^{-h_K}) \f{1}{P_{13}^2} A(\hat K^{h_K}, \hat 2, \dots, n),
\label{special1}
\ee
and,
\be
\sum_{h_K} A(\hat 1,\dots,-\hat K^{-h_K}) \f{1}{P_{23}^2} A^{(a)}(\hat K^{h_K}, \hat 2, 3).
\label{special2}
\ee
The contribution \eqref{special1} has a net $\frac{1}{\ap{13}}$ in the
explicit propagator, but the 3-point amplitude is proportional to
$\ap{13}^s$ so \eqref{special1} is not singular for $s\geq 1$.  The
term \eqref{special2} has potentially singular behavior because the
shifted momenta,
\be
\rr{\hat 2} = \frac{\ap{12}}{\ap{13}} \rr{\hat 3} \quad,
\rr{\hat K}|\hat K] = 
\left(-\frac{\ap{12}}{\ap{13}} \rr{3} \right)
\left(|2] + {\ap{13}}{\ap{12}} |3]\right),
\label{eqn:kinematics}
\ee
go to infinity as
$\ap{13}\toZ$.  The growing momentum is suggestive of a large-$z$ BCFW
shift -- a limit in which we expect lower-point amplitudes to approach
zero.  To make this explicit, we write the momenta $\hat 1$ and $-\hat
K$ in a somewhat unusual way.  We first define
 ``initial'' momenta,
\be
\rr{1_0} |1_0] = \rr{1} \left( |1] - \f{\ap{23}}{\ap{12}}
|3] \right) 
\quad 
\rr{K_0} |K_0] = \rr{2} \left( |2] + \f{\ap{13}}{\ap{12}}
|3] \right) ;
\ee
note that as $\ap{13}\toZ$, these momenta approach limits $\rr{1_0}
|1_0] \rightarrow \rr{1_*} |1_*] = K_{13}$ and $\rr{K_0} |K_0]
      \rightarrow \rr{K_*} |K_*] = p_2$. 
The hatted momenta in \eqref{special2} can be written as a BCFW shift
of the ``initial'' momenta:
\be
\hat p_1 = \rr{1} ( |1_0] - z_* |K_0]),\;
-\hat K = ( \rr{K_0} - z_* \rr{1_0}) |K_0],
\quad \mbox{ where } z_* = \frac{\ap{23}}{\ap{13}}.
\ee
All invariants involving ``starred'' momenta are generically finite,
and differ from invariants of ``naught'' only by terms supressed by
$1/\ap{13}$.  Therefore, the scaling of the $(n-1)$-point amplitude in
\eqref{special2} as $\ap{13} \toZ$ is dictated by the large-$z$
scaling of
\be
A_{BCF,\sh{1_*,K_*}}(1_*,K_*,4,\dots,n),
\ee
which we derive for spin-1 in Appendix \ref{app:scaling}.  Using this
result and the explicit kinematics of \eqref{eqn:kinematics} for the
three-point amplitude, we find that, as long as $\sh{1,2}$ is a 
``valid'' BCFW shift, \eqref{special2} is non-singular as
$\ap{13}\toZ$.  This is summarized in Table \ref{tab:checkGauge}.  We
highlight three examples:
\begin{itemize}
\item If $h_1 = h_2 = +1$ and $h_3 = +1$, then the only non-zero term comes
from $h_K=-1$. The three-point amplitude diverges as 
$1/\ap{13}$, but the scaling of the $(n-1)$-point amplitude under a
large $\sh{+,+}$ shift is $1/z_* \sim \ap{13}$, so the product has a finite
limit.  
\item If instead $h_3=-1$, then the 3-point amplitude
vanishes as $\ap{13}^3$, but the $(n-1)$-point amplitude
grows as $z_*^3$ under a $\sh{+,-}$ shift; again the product
approaches a finite limit as $\ap{13}\toZ$.  
\item For ``invalid'' BCFW shifts $\sh{+,-}$, term (c) does contribute
  a singularity (in fact, an unphysical multiple pole!).  For example,
  we consider the contribution from $h_K = +1$ when $h_1 = h_3 = +1$
  and $h_2 = -1$.  The three-point amplitude in \eqref{special2}
  scales as $1/\ap{13}$ and the $(n-1)$-point amplitude scales as
  $z_*^3$ under the $\sh{1^+,(-K)^{-}}$ shift -- as $\ap{13}\toZ$, the
  product can have an unphysical $\ap{13}^{-4}$ singularity!
\end{itemize}

\begin{table}[htbp]\begin{center}
\begin{tabular}{|ccc|c|cc|c|}
\hline
$h_1$ & $h_2$ & $h_3$ & $h_K$ & $(n-1)$-point & Three-point & Total scaling \\ 
\hline
+ & + & + & - & $\epsilon$ & $1/\epsilon$ & $1$  \\
+ & + & - & + & $1/\epsilon^3$ & $\epsilon^3$ & $1$ \\
\hline
- & + & + & - & $\epsilon$ & $1/\epsilon$ & $1$ \\
- & + & - & + & $\epsilon$ & $\epsilon^3$ & $\epsilon^4$ \\
\hline
- & - & + & + & $\epsilon$ & $1/\epsilon$ & $1$ \\
- & - & - & X &\multicolumn{3}{c|}{(c) vanishes identically}\\
\hline\hline
+ & - & + & + & $1/\epsilon^3$ & $1/\epsilon$ & $\epsilon^{-4}$\\
+ & - & - & X & \multicolumn{3}{c|}{(c) vanishes identically}\\
\hline
\end{tabular}\end{center}
\caption{\label{tab:checkGauge}Scaling of the term \eqref{special2}
  with $\epsilon \equiv \ap{13}$ as $\ap{13}\toZ$.  For given $h_2$
  and $h_3$, at most one $h_K$ has a non-vanishing contribution to
  this diagram.  An ``X'' denotes helicity combinations in which the
  3-point amplitude $A^{(a)}(2,3,K)$ vanishes for all $h_K$.  We
  evaluate the 3-point scaling explicitly using the kinematics of
  \eqref{eqn:kinematics} and the $(n-1)$-point scaling using $z \sim
  1/\epsilon$ and the general scaling found in Appendix
  \ref{app:scaling}.  The bottom two lines correspond to the
  ``invalid'' shift. }

\end{table}

The scaling of gravity amplitudes is known to be the square of the
gauge theory scaling described above -- this scaling suffices to
guarantee that the term \eqref{special2} does not contribute to the
factorization limit in gravity, either.  However, unlike gauge theory,
we have not found an S-matrix argument for why this must be so.  The
pivotal role of  $1/z^2$ scaling in showing that gravity amplitudes
factorize is striking and unexpected --- this is a faster scaling than
is required to prove BCFW in field theory, for example!  It is
possible that in some cases, the boundary-term of \eqref{eqn:bdry}
cancels the one arising from \eqref{special2}.   But such a
cancellation is unlikely to be universal. For example, if $h_1 = h_2 =
+1$ and $h_3 = -1$, the boundary term of \eqref{eqn:bdry} is
associated with a coefficient of $1/z$ in a falling amplitude,
while \eqref{special2} is the product of a \emph{growing}
$(n-1)$-point amplitude with a 3-point amplitude that falls as $\ap{13}^6$!

%%%%%%%%%%%%%%%%%%%%%%%%%%%%%%%%%%%%%%%
\section{Summary and Remarks}
The BCFW recursion relation is a remarkable formula, demonstrating
that the entire structure of Yang-Mills and gravity amplitudes at
tree-level can be derived recursively from 3-point amplitudes, which
are fully specified by their Lorentz structure.  The coefficients of
these 3-point amplitudes must satisfy additional consistency
conditions, identified in \cite{Benincasa:2007xk}, for the BCFW
four-point amplitude to be well-defined. 

We have shown that in fact, when these consistency conditions are
violated, no four-particle amplitude with correct Lorentz
transformation \emph{and} factorization properties can be defined.  Thus, the
consistency requirements are quite strong --- their violation truly
indicates an inconsistency of the set of three-point amplitudes, not
merely of a BCFW construction.  We have used it to reproduce several
standard results (Jacobi identity, charge conservation, and
equivalence principle) for spins 1 and 2 coupled to other particles of
the same spin, or to scalar matter.

Related to this four-particle factorization requirement is a necessary
criterion for BCFW constructions: when a four-point amplitude exists,
we can ask whether a given BCFW construction reproduces it.  We find
that the constructions known to be valid from gauge theory (shifts
$\sh{--}$, $\sh{-+}$, and $\sh{++}$) do, while attempts to generate
amplitudes using a wrong-helicity BCF shift ($\sh{+-}$) result in
``amplitudes'' with unphysical multiple poles.  Thus, S-matrix
consistency alone shows that these shifts are invalid.

The self-consistency of BCFW amplitudes beyond four-point is not
obviously guaranteed by these four-particle conditions.  We have
demonstrated that the four-particle conditions are in fact sufficient,
for interacting spin-1 fields, by showing inductively that any BCFW
construction using a valid shift has the correct factorization
properties in all complex-momentum factorization limits.  The
equivalence of different BCFW constructions of the same amplitude (the
generalization of the four-particle test of \cite{Benincasa:2007xk})
follows from this fact and power-counting.  The factorization
arguments fail for the invalid $\sh{+-}$ shifts.  We have also
outlined a proof of the analogous statement for gravity.  The argument
requires that $n$-point amplitudes scale at large-$z$ BCFW shifts in
the same way as the fundamental three-point amplitudes, i.e. as
$z^{3s}$  for $\sh{+-}$ shifts and $1/z^{s}$ for all others.  We have
proved this result for spin-1 (but not yet for spin-2), again using
only S-matrix arguments.

One may ask why BCFW amplitudes should
include all possible poles, given that they are only explicitly
constructed by factorizing diagrams on a subset of poles.  Formally,
the sufficiency of considering this subset is guaranteed by the
large-$z$ behavior of amplitudes; the physical intuition suggested by
the study of four-particle amplitudes is that BCFW works precisely
when the helicity structure of amplitudes requires them to have
simultaneous singularities in multiple invariants (e.g. $1/(stu)$ in
gravity four-point amplitudes).  

These two very different explanations of BCFW may be more related than
they appear --- we have argued that the large-$z$ scaling of spin-1
amplitudes can be derived from their factorization properties, mass
dimension, and transformation under helicity rotations of the two
shifted legs.  Requiring that amplitudes factorize drastically
constrains the functional form of any part of an amplitude that grows
faster (or falls slower) with $z$ than the three-particle amplitudes.  In
fact, one cannot write down any amplitude that grows faster with $z$
than three-particle amplitudes, factorizes, and transforms properly under
helicity rotations of the two shifted legs.  This argument may
generalize to spin-2, and is the closest analogue we are aware of in
the spinor-helicity language to the ``spin Lorentz'' symmetries found
in the background-field approach of \cite{ArkaniHamed:2008yf}.

The results of this paper are completely unsurprising --- indeed, the
factorization of gauge and gravity amplitudes is a consequence of
their equivalence to well-known field theories.  However, the
technical mechanisms for achieving factorization are somewhat
remarkable, and suggestive.  First, it is striking that the conditions
appear as factorization conditions \emph{only at complex momenta}
(e.g. the constraints on four-particle gauge theory amplitudes can
only be exposed by considering the singularity as $[12]\toZ$ with
$\ap{12}$ finite -- if both go to zero simultaneously, the amplitude
is non-singular).  Moreover, the BCFW terms
that are singular as $[12]\toZ$ are precisely the
\emph{soft photon/graviton} singularities that appear in Weinberg's
classic derivations of charge conservation and the equivalence
principle --- even though the limit we consider need only be collinear
(in one spinor factor of the momenta).  

Particularly noteworthy is the connection between the requirement of
factorization in ``wrong-helicity'' two-particle poles that are not
exposed by BCF (e.g. $\ap{13}\toZ$ when $|1]$ is BCF-shifted) and the
$1/z^2$ scaling of gravity amplitudes.  The sensitivity of this
factorization limit to the  $1/z$ coefficient of gravity
amplitudes is so striking because the field theory argument for BCFW
does not require this coefficient to vanish -- in that construction,
it is a seemingly irrelevant accident.  In contrast,
the appearance of $1/z^2$ scaling in the factorization requirement
suggests a connection between the (hard) large-$z$ behavior of gravity
amplitudes and factorization in collinear limits.  
We should, however, point out that the derivation
in Section \ref{sec:wrongfac} includes \textbf{two} appeearances of the $1/z$
coefficient --- it is conceivable that in some consistent
theories, these coefficients do not vanish but one cancels the other.

The structure found in these familiar theories suggests three related
directions at tree-level for further investigation.  First, it is
possible that the ``constructive'' approach used here could be used to
find new BCFW recursive constructions (presumably equivalent to known
Lagrangian theories). There are sets of three-point amplitudes for which a consistent four-particle amplitude
exists, but cannot be obtained by any BCFW recursion relation
(e.g. $\phi^3$ theory, or interacting spin-1 particles with non-zero
(+,+,+) and (-,-,-) vertices).  It seems likely that a more general
recursion relation exists, that allows us to generate an S-matrix for
any such set of primitive amplitudes. We also expect that analogous structural constraints (such as anomalies) appear at one-loop. 
Finally, in higher dimensions, there are theories with no known Lorentz- and
gauge-invariant action, and it is possible that generalizations of our
construction to higher dimensions would permit the study of an
S-matrix for such theories.  

%%%%%%%%%%%%%%%%%%%%%%%%%%%%%%%%%%%%%%%
\section*{Acknowledgments}
We are grateful to Nima Arkani-Hamed, Clifford Cheung, Jared Kaplan, and 
Michael Peskin for valuable discussions and feedback on this work.

%%%%%%%%%%%%%%%%%%%%%%%%%%%%%%%%%%%%%%%
\appendix
\section{Large-$z$ Scaling of $n$-point Gauge Theory Amplitudes}
\label{app:scaling}
The large-$z$ scaling behavior of gauge theory amplitudes
$A^{(n)}(p_1(z),p_2(z),\dots,n)$, when two legs are deformed by a BCFW
shift $\sh{1,2}$ parametrized by $z$
\be
|1] \rightarrow |1] + z |2], \quad \rr{2} \rightarrow \rr{2} - z
      \rr{1},
\label{12LargeZ}
\ee
is known to be determined by the
helicities of legs 1 and 2.  For $h_1=+1$ and $h_2 = -1$, $A^{(n)}$
grows with $z^3$ at large $z$, and for other helicity choices (+/+,
-/-, and -+) it falls as $1/z$.  

In field theory, the former growth follows from naive power-counting,
whereas the $1/z$ behavior at large $z$ is not obvious from
diagrammatic arguments.  For BCFW amplitudes, these scalings are
guaranteed by the factorization and Lorentz structure of the
amplitudes, as we will now show.  
We use the $(n-1)$-point scaling result (which depends only on
factorization of $(n-1)$-point amplitudes) in the $n$-point
factorization argument of \ref{sec:wrongfac}.

We begin with the $1/z$ scalings under ``valid'' BCF shifts. Of
course, this scaling is guaranteed by the BCFW construction (in which
every term has a pole $\sim 1/(z-z_*)$), if
$A^{(n)}$ was generated by the same BCFW shift $\sh{1,2}$.  

In fact, this is all we need, because all valid BCFW shifts must
generate the same $n$-point amplitude.  We have shown in Section
\ref{sec:4} that any two BCFW amplitudes have identical singularity
structure at every kinematic singularity; therefore, they can only
differ by completely non-singular terms.  However, they can depend
only on the dimensionless coupling constants of the three-point
amplitudes so the total mass dimension of any such term would be
non-negative.  Power-counting alone suffices to rule out such terms in
5-point amplitudes and higher (even at four-point, power-counting
and correct helicity transformation properties prohibit such new
terms).  

Therefore, all BCFW amplitudes agree, and must fall as $1/z$ in
any limit that corresponds to a ``valid'' shift.

\paragraph{A Pure Scaling/Factorization Argument}
We could have obtained the same result by a more general inductive
argument that does not rely on the functional form of the BCFW
amplitudes but \emph{only} on factorization and power-counting.
Instead, we assume that lower-point amplitudes have the correct
scaling as the $z$-parameter of \eqref{12LargeZ} approaches infinity,
and proceed by induction.  This argument will also apply to the $z^3$
scaling of ``wrong'' shifts.

Suppose there is some component $f^0(p_1(z),p_2(z),\dots,p_n)$ of an
$n$-point amplitude that scales as $z^0$.  It can necessarily be built
out of $z$-independent invariants \be
\ap{12},\,[12],\,[XY],\,P_{XY..Z}^2
\label{eqn:basicZindep}
\ee 
and
\be
\ap{1X},\,[2X],
\ee
or combinations of invariants
\be
\frac{[1X]}{[1X]},\,\frac{[1X]}{\ap{2X}},
\label{eqn:Zindepratios}
\ee
where $X,Y$ denotes any legs besides 1 and 2 (in expressions with
multiple $X$'s, they should be regarded as distinct, arbitrary legs).

However, the form of $f^0$ is tightly constrained by complex
factorization --- it cannot have poles in the invariants of
\eqref{eqn:basicZindep}.  For example, at small $[XY]$ we have the
factorization: \be \lim_{[XY]\toZ} [XY]\ap{XY}
A^{(n)}(p_1(z),p_2(z),\dots,n) = A^{(n-1)}(p_1(z),p_2(z),...,K)
A^{(h)}(K,X,Y).  \ee On the right-hand side, the $(n-1)$-point
amplitude scales as $1/z$ by induction and the three-point amplitude
is manifestly $z$-independent; therefore all terms in $A^{(n)}$
singular as $[XY]\toZ$ must also scale as $1/z$.  Identical logic
applies to the other poles in \eqref{eqn:basicZindep}.

We now wish to show that any function $f^0$ that satisfies the
$1$ and $2$ helicity transformation properties for an amplitude, is
$z$-independent, and has no poles in the invariants of \eqref{eqn:basicZindep}
must have non-negative mass dimension.  To begin, we write down
particular solutions with proper helicity transformations:
\be
a_{-,-} = \frac{\ap{1X}^2}{[2X]^2}\,(d=0),\quad
a_{+,+} = \frac{[2X]^2}{\ap{1X}^2}\,(d=0),\quad
a_{-,+} = \ap{1X}^2[2X]^2 (d=4).
\ee
General solutions can be obtained by multiplying these by
$z$-independent, helicity-scalar invariants that have no
  forbidden poles:
\be
\frac{[1X]}{[1Y]},\quad
\frac{\ap{1X}}{\ap{1Y}},\quad
\frac{[2X]}{[2Y]},\quad
\frac{\ap{2X}}{\ap{2Y}},\quad
\frac{\ap{2X}[2X]}{\ap{1X'}[1Y']},\quad
\ap{12}[12],\quad
\frac{[12]\ap{1X}}{[2Y]},\,
\frac{\ap{12}[2X]}{\ap{1Y}}\quad
[XY],\,\ap{XY}.
\ee
but all such terms have non-negative mass dimension.  

Five-point and higher amplitudes must have negative mass dimension,
which we can only obtain by re-introducing ``forbidden'' poles or
putting a negative power of $z$ in the denominator.  For $\sh{+,+}$
and $\sh{-,-}$ shifts, pure power-counting is not sufficient to rule
out amplitudes that scale as $z^0$, but we have already constructed
the unique four-particle gauge amplitudes, and they scale as $1/z$
(the additional constraint in this case comes from helicity
transformations under the two unshifted legs).  

To summarize: if $k$-point amplitudes scale as $1/z$ under large BCF
shifts for k<n, any term in an $n$-point amplitude that violates this
scaling must not have singularities on which it factorizes into a
$z$-independent amplitude and a lower-point amplitude with $z$-scaling
determined by the induction hypothesis.  But we cannot write any
function consistent with factorization and helicity transformation
that has the correct mass dimension to appear in an $n$-point
amplitude but has none of the forbidden singularities.  Therefore, the
$n$-point gauge amplitudes must fall as fast as their lower-point
counterparts, namely as $1/z$.

\paragraph{$z^3$ Growth}
For helicities $1^+$ and $2^-$, we proceed as in the previous
argument, but in this case we attempt to construct a function $f^4$
that scales as $z^4$, transforms correctly under helicity rotations of
$1$ and $2$, and has negative mass dimension but no forbidden
singularities.

A particular solution to the first two requirements is
\be
a_{+,-} = [1X]^2\ap{2Y}^2\,(d=4).
\ee
Again, however, we cannot obtain any term of dimension 0 or lower
without introducing forbidden poles.  Therefore, the leading behavior
of $n$-point amplitudes must be, as in the 3-point case, $z^3$.

\section{Absence of Derivative Contributions to $[12]$ Singularity in
Gravity}\label{app:derivatives}
In this appendix, we revisit the expression \eqref{eqn:iterm} for the
singularity as $[12]\toZ$ in gravity.  The three-point amplitudes in
\eqref{eqn:iterm} diverge
as $[12]^{-2}$ for gravity, so $O([12])$ corrections to the $(n-1)$-point
amplitudes in \eqref{eqn:iterm} will contribute to the singularity unless
they cancel when summed over all legs.  We consider these contributions
here (all other terms of order $1$ or $[12]$ were considered in
Sec.~\ref{sec:12pole}).  

We point out a property of \eqref{eqn:iterm} that we will use
repeatedly: the $[12]^{-2}$ singularity in the $i$'th term is
proportional to $[2i]\ap{1i}$.  Therefore, any contribution of
$O([12])$ that is independent of $i$ will appear as 
\be
c \left(\sum_{i} [2i] \ap{1i}\right) = 0 \label{momentumRule}
\ee
by momentum conservation.  Here, we are considering $O([12])$ effects,
so only those that are $i$-dependent can contribute.

To study the differences in the amplitude at finite $[12]$, we should
construct a sufficiently general explicit path in $n$-particle
kinematics, parameterized by $\epsilon$, with $[12] \propto \epsilon$
at small $\epsilon$.  On any such path, at least one leg besides 1 and
2 will have to shift momentum by $O(\epsilon)$ (the only
momentum-conserving deformations that affect only legs 1 and 2,
but keep these legs null and conserve momentum are BCF shifts, and
these do not change $[12]$).  The amplitudes will depend on this
deformation, \emph{but the terms for different $i$ will all depend in
  the same way}, so when summed, they do not contribute to the singularity by
\eqref{momentumRule}.  

This allows us to calculate using a particularly simple path, for
instance one in which only leg $n$ compensates for the deformations of
legs 1 and 2:
\be
\rr{1}=(1-\alpha)\rr{\lambda},\quad \rr{2}=\alpha \rr{\lambda} - 
\epsilon\rr{\mu},\, |2] =
  |n],\quad \mbox{and} \rr{n(\epsilon)} = \rr{n}-\epsilon\rr{\mu},
\ee
for arbitrary $\rr{\mu}$, $\rr{\lambda}$, and $\alpha$.  
The BCFW shift in the $i$'th term will deform $|1]$, and replace $|i]$
    with a slightly shifted momentum $\hat K_i$ --- specifically, it
    will involve an $(n-1)$-point amplitude 
\be
A\left(\rr{1}\left(|1] + \epsilon \fap{i\mu}{i1} |n]\right),3,\dots,
\rr{i} \left(|i] + \epsilon \fap{1\mu}{1i}|n] \right), \dots,
\left(\rr{n} - \epsilon\rr{\mu}\right)|n]\right).
\ee
Then 
\be
\f{dA_i}{d\epsilon} = 
\f{\ap{i\mu}}{\ap{i1}} 
\left( \tilde \lambda_n \cdot \frac{\partial A}{\partial \tilde \lambda_1}  
 \right) 
+ 
\fap{1\mu}{1i} \left(\tilde \lambda_n \cdot \f{\partial A}{\partial
  \tilde \lambda_i}\right) 
+ \mbox{ $i$-indep. terms}.\label{i_derivative}
\ee
The sum in \eqref{eqn:iterm} contains a term 
\be
\f{[1n]}{\alpha \ap{1\mu}} \sum_{i \neq 1,n} [ni]\ap{1i} \f{dA_i}{d\epsilon}.
\ee
Inserting just the first term of \eqref{i_derivative} into this sum,
we recover
\be
-\left(\tilde \lambda_n \cdot \f{\partial A}{\partial
  \tilde \lambda_1}\right) \sum_{i \neq 1,n} \ap{i\mu}[ni] 
= \left(\tilde \lambda_n \cdot \f{\partial A}{\partial
  \tilde \lambda_1}\right) \times \ap{1\mu} [n1],\label{1_summand}
\ee
which has the same form as
the second term, but for $i=1$.  Thus the total is a sum (which can
now be written over all legs, since the $i=n$ term vanishes and the
$i=1$ term comes from \eqref{1_summand}):
\be
\ap{1\mu} \sum_i  [ni] \left(\tilde \lambda_n \cdot \f{\partial A}{\partial
  \tilde \lambda_j}\right).
\ee
This is identically zero on any bi-spinor $[xy]$ involving two of the
legs that are summed over, since it receives opposite contributions
from $i=x$ and $i=y$.  Hence it also vanishes on any arbitrary function
of kinematic invariants built out of the legs.  Thus, we are justified
in ignoring this source of $[12]$-dependence in the factorization
argument of Sec.~\ref{spin2_12}.

% Consider any
%deformation of legs as a function of epsilon, such that [12]->0 at
%small epsilon, and momentum is conserved for all epsilon, and
%consider epsilon->0 limit.  For example, the one where only $n$
%compensates.  Note that at least one leg will have to shift momentum,
%but these are j-independent.  Note terms that are
%j-independent produce no singularity ---  summing over all legs the
%[2j]<2j> we'll get a factor of [12]<12>; this factor and the o([12])
%from the fact that this is the *derivative* of the A wrt epsilon
%cancel the singularity.  The two j-dependent terms are delta|2] =
%... and delta | j] = ....  A few lines of manipulation to get a sum
%over all legs of the sub-amplitude, then kinematics.  Should take < 1 page.

%%%%%%%%%%%%%%%%%%%%%%%%%%%%%%%%%%%%%%%%%%%%%%%%
%% You may have to change the BibTeX style below, depending on your
%% setup or preferences.
%%
%%
%% For The AIP proceedings layouts use either
%%%%%%%%%%%%%%%%%%%%%%%%%%%%%%%%%%%%%%%%%%%%

\bibliographystyle{JHEP}   % if natbib is available
%\bibliographystyle{aipproc}   % if natbib is available
%\bibliographystyle{aipprocl} % if natbib is missing

%%%%%%%%%%%%%%%%%%%%%%%%%%%%%%%%%%%%%%%%%%%
%% You probably want to use your own bibtex database here
%%%%%%%%%%%%%%%%%%%%%%%%%%%%%%%%%%%%%%%%%%%
\bibliography{CompFactPaper}

\providecommand{\href}[2]{#2}\begingroup\raggedright\begin{thebibliography}{10}

\bibitem{Weinberg:1964ev}
S.~Weinberg, {\it {Feynman Rules for Any Spin. 2. Massless Particles}},  {\em
  Phys. Rev.} {\bf 134} (1964) B882--B896.

\bibitem{Weinberg:1964ew}
S.~Weinberg, {\it {Photons and Gravitons in s Matrix Theory: Derivation of
  Charge Conservation and Equality of Gravitational and Inertial Mass}},  {\em
  Phys. Rev.} {\bf 135} (1964) B1049--B1056.

\bibitem{Weinberg:1965rz}
S.~Weinberg, {\it {Photons and gravitons in perturbation theory: Derivation of
  Maxwell's and Einstein's equations}},  {\em Phys. Rev.} {\bf 138} (1965)
  B988--B1002.

\bibitem{Britto:2004ap}
R.~Britto, F.~Cachazo, and B.~Feng, {\it {New recursion relations for tree
  amplitudes of gluons}},  {\em Nucl. Phys.} {\bf B715} (2005) 499--522,
  [\href{http://xxx.lanl.gov/abs/hep-th/0412308}{{\tt hep-th/0412308}}].

\bibitem{Britto:2005fq}
R.~Britto, F.~Cachazo, B.~Feng, and E.~Witten, {\it {Direct proof of tree-level
  recursion relation in Yang- Mills theory}},  {\em Phys. Rev. Lett.} {\bf 94}
  (2005) 181602, [\href{http://xxx.lanl.gov/abs/hep-th/0501052}{{\tt
  hep-th/0501052}}].

\bibitem{Bedford:2005yy}
J.~Bedford, A.~Brandhuber, B.~J. Spence, and G.~Travaglini, {\it {A recursion
  relation for gravity amplitudes}},  {\em Nucl. Phys.} {\bf B721} (2005)
  98--110, [\href{http://xxx.lanl.gov/abs/hep-th/0502146}{{\tt
  hep-th/0502146}}].

\bibitem{Cachazo:2005ca}
F.~Cachazo and P.~Svrcek, {\it {Tree level recursion relations in general
  relativity}},  \href{http://xxx.lanl.gov/abs/hep-th/0502160}{{\tt
  hep-th/0502160}}.

\bibitem{Benincasa:2007qj}
P.~Benincasa, C.~Boucher-Veronneau, and F.~Cachazo, {\it {Taming tree
  amplitudes in general relativity}},  {\em JHEP} {\bf 11} (2007) 057,
  [\href{http://xxx.lanl.gov/abs/hep-th/0702032}{{\tt hep-th/0702032}}].

\bibitem{ArkaniHamed:2008yf}
N.~Arkani-Hamed and J.~Kaplan, {\it {On Tree Amplitudes in Gauge Theory and
  Gravity}},  {\em JHEP} {\bf 04} (2008) 076,
  [\href{http://xxx.lanl.gov/abs/0801.2385}{{\tt 0801.2385}}].

\bibitem{Benincasa:2007xk}
P.~Benincasa and F.~Cachazo, {\it {Consistency Conditions on the S-Matrix of
  Massless Particles}},  \href{http://xxx.lanl.gov/abs/0705.4305}{{\tt
  0705.4305}}.

\bibitem{SongHe}
S.~He, {\it {Consistency Conditions on S-Matrix of Spin 1 Massless Particles}},
   \href{http://xxx.lanl.gov/abs/0811.3210}{{\tt 0811.3210}}.

\bibitem{Witten:2003nn}
E.~Witten, {\it {Perturbative gauge theory as a string theory in twistor
  space}},  {\em Commun. Math. Phys.} {\bf 252} (2004) 189--258,
  [\href{http://xxx.lanl.gov/abs/hep-th/0312171}{{\tt hep-th/0312171}}].

\bibitem{Berends:1981rb}
F.~A. Berends, R.~Kleiss, P.~De~Causmaecker, R.~Gastmans, and T.~T. Wu, {\it
  {Single Bremsstrahlung Processes in Gauge Theories}},  {\em Phys. Lett.} {\bf
  B103} (1981) 124.

\bibitem{DeCausmaecker:1981bg}
P.~De~Causmaecker, R.~Gastmans, W.~Troost, and T.~T. Wu, {\it {Multiple
  Bremsstrahlung in Gauge Theories at High- Energies. 1. General Formalism for
  Quantum Electrodynamics}},  {\em Nucl. Phys.} {\bf B206} (1982) 53.

\bibitem{Kleiss:1985yh}
R.~Kleiss and W.~J. Stirling, {\it {Spinor Techniques for Calculating p anti-p
  $\to$ W+- / Z0 + Jets}},  {\em Nucl. Phys.} {\bf B262} (1985) 235--262.

\bibitem{Dixon:1996wi}
L.~J. Dixon, {\it {Calculating scattering amplitudes efficiently}},
  \href{http://xxx.lanl.gov/abs/hep-ph/9601359}{{\tt hep-ph/9601359}}.

\bibitem{Bern:2007dw}
Z.~Bern, L.~J. Dixon, and D.~A. Kosower, {\it {On-Shell Methods in Perturbative
  QCD}},  {\em Annals Phys.} {\bf 322} (2007) 1587--1634,
  [\href{http://xxx.lanl.gov/abs/0704.2798}{{\tt 0704.2798}}].

\bibitem{Cheung:2008dn}
C.~Cheung, {\it {On-Shell Recursion Relations for Generic Theories}},
  \href{http://xxx.lanl.gov/abs/0808.0504}{{\tt 0808.0504}}.

\bibitem{Berends:1987cv}
F.~A. Berends and W.~Giele, {\it {The Six Gluon Process as an Example of
  Weyl-Van Der Waerden Spinor Calculus}},  {\em Nucl. Phys.} {\bf B294} (1987)
  700.

\bibitem{Mangano:1987xk}
M.~L. Mangano, S.~J. Parke, and Z.~Xu, {\it {Duality and Multi - Gluon
  Scattering}},  {\em Nucl. Phys.} {\bf B298} (1988) 653.

\bibitem{Mangano:1988kk}
M.~L. Mangano, {\it {The Color Structure of Gluon Emission}},  {\em Nucl.
  Phys.} {\bf B309} (1988) 461.

\end{thebibliography}\endgroup

%%%%%%%%%%%%%%%%%%%%%%%%%%%%%%%%%%%%%%%%%%%
%% Just a reminder that you may have to run bibtex
%% All of it up to \end{document} can be removed
%% if you don't like the warning.
%%%%%%%%%%%%%%%%%%%%%%%%%%%%%%%%%%%%%%%%%%%
\IfFileExists{\jobname.bbl}{}
 {\typeout{}
  \typeout{******************************************}
  \typeout{** Please run "bibtex \jobname" to optain}
  \typeout{** the bibliography and then re-run LaTeX}
  \typeout{** twice to fix the references!}
  \typeout{******************************************}
  \typeout{}
 }
%%%%%%%%%%%%%%%%%%%%%%%%%%%%%%%%%%%%%%%%%%%
%%%%%%%%%%%%%%%%%%%%%%%%%%%%%%%%%%%%%%%%%%%
\end{document}